\pdfoutput=1
\documentclass[aps,pre,reprint,longbibliography,nofootinbib,superscriptaddress]{revtex4-2}
\usepackage{amsmath}
\usepackage{amssymb}
\usepackage{amsfonts}
\usepackage{mathtools}
\usepackage{graphicx}
\usepackage{float}
\usepackage{placeins}
\usepackage{booktabs}
\usepackage{multirow}
\usepackage{algorithm}
\usepackage{algpseudocode}
\usepackage[table]{xcolor}
\usepackage{subcaption}
\usepackage{hyperref}
\hypersetup{colorlinks=true, allcolors=blue!60!black}

\newcommand{\dysts}{\texttt{dysts}}
\newcommand{\teps}{\widetilde\varepsilon}
\newcommand{\drel}{\delta_{\mathrm{rel}}}
\newcommand{\Nens}{N_{\mathrm{ens}}}
\newcommand{\gorc}{\gamma_{\mathrm{orc}}}

\begin{document}

\title{Partial Multistep Prediction of Chaotic Time Series: What Remains Predictable beyond the Horizon of Predictability?}
\author{Vasilii A. Gromov}
\author{Maksim Lomakin}
\author{Aleksei Kudryashov}
\author{Korney K. Tomashchuk}
\affiliation{HSE University}

\begin{abstract}
Sensitive dependence on initial conditions limits long-horizon prediction of chaotic systems even for deterministic dynamics, yet some dynamical information may remain recoverable after precise state information is lost. Recursive multistep methods are fragile because they reuse each intermediate prediction as input, allowing local errors to propagate through the rollout. We investigate whether partial multistep prediction can recover predictive information beyond the predictability horizon by omitting intermediate positions instead of assigning them unreliable values. The method uses a lightweight pattern-based local predictor. We analyze partial trajectories generated by different omission patterns and quantify the trade-off between accuracy on retained positions and the fraction classified as non-predictable. Experiments on the Lorenz system show that admissible masks contain trajectories that remain accurate far beyond the horizon of predictability, and an agreement-based ensemble recovers part of this information without future observations at the cost of a growing rejected fraction. At short horizons a broad accepted subset retains accuracy at moderate rejection, whereas at long horizons positive geometric skill requires increasingly strong selection. In this regime retained predictions no longer localize the state within a Lorenz scroll but still identify the occupied scroll, so topological structure outlasts geometric accuracy. A noise-based control reproduces this survival only after tuning of its perturbation scale, unlike masking. Across a heterogeneous collection of chaotic systems the ensemble strictly extends the positive-$R^2$ horizon of the recursive baseline for every retained system.
\end{abstract}
\maketitle

\section{Introduction}
\label{sec:introduction}

This paper studies partial multistep prediction of chaotic time series: a construction that omits the intermediate positions it cannot predict reliably, so that some structure can be recovered beyond the horizon of predictability. For chaotic dynamics, small initial and model errors tend to restrain accurate long-horizon prediction. The separation of nearby trajectories scales as
\begin{equation}
\delta(\tau)\sim \delta_0 e^{\lambda_{\max}\tau},
\end{equation}
where $\lambda_{\max}>0$ is the largest Lyapunov exponent. For an admissible error level $\delta_{\max}$, this relation gives the estimate
\begin{equation}\label{eq:tpred}
T_{\mathrm{pred}} \sim \lambda_{\max}^{-1} \log\!\left(\frac{\delta_{\max}}{\delta_0}\right)
\end{equation}
for the horizon of predictability \cite{MalinetskyPotapov2000,Boffetta2002Predictability,KantzSchreiber2004}. Empirical error growth need not follow a single exponential law over all scales, but the same instability makes recursive multistep prediction fragile. An inaccurate intermediate value becomes part of the input at later lead times and can redirect the predicted trajectory.

\begin{figure}[tbp]
  \centering
  \includegraphics[width=\linewidth]
  {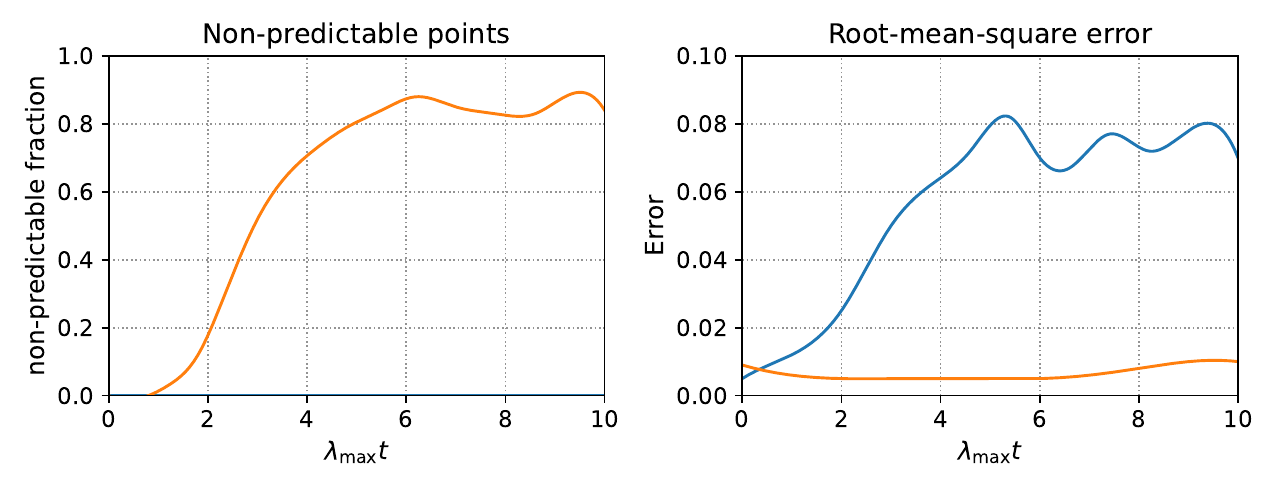}
  \caption{Schematic illustration of the target behaviour of partial multistep prediction. Panel (a) gives the fraction of non-predictable intermediate positions as a function of the prediction horizon $h$; panel (b) gives RMSE on the positions for which the method produces a prediction. The blue curves correspond to recursive prediction without non-predictable intermediate positions, for which the error increases with the horizon. The orange curves illustrate partial multistep prediction. The number of non-predictable positions increases, while the error on the remaining predictable positions stays small and approximately constant.}
  \label{fig:desirable_non_pred_vs_error}
\end{figure}

Most forecasting methods nevertheless use a recursive baseline that assigns a value to every intermediate position. Partial multistep prediction takes a different route. Building on the non-predictable observations of predictive-clustering approaches~\cite{GromovBorisenko}, Gromov and Baranov allow the recursive construction to skip selected intermediate positions~\cite{GromovBaranov}, declaring them non-predictable instead of assigning them unreliable values. Prediction quality then rests on two competing criteria: accuracy on the retained positions and the fraction of positions classified as non-predictable. Figure~\ref{fig:desirable_non_pred_vs_error} illustrates the target limiting behaviour. The error of the recursive baseline grows with the horizon, whereas partial prediction concedes an increasing fraction of non-predictable positions to keep the error on the retained ones small.

Earlier work established that omitting selected intermediate values can prevent individual prediction errors from contaminating the entire rollout, but it characterizes neither the family of partial continuations generated by different omission patterns nor the limits of what such continuations can recover. This paper asks the resulting question directly: what remains predictable beyond the horizon of predictability of a chaotic time series, and at what cost? Partial multistep prediction makes the question quantitative. Conditional accuracy on the retained positions measures what remains, and the non-predictable fraction measures the cost. We trace this trade-off over lead times.

Three aspects of this question structure the study. The first is existence and extraction: whether the admissible mask class contains long low-error partial continuations at all, and how much of their accuracy survives when a selection procedure cannot access the future. The second is the nature of the surviving information: whether accepted predictions still localize the state on the attractor at long lead times, or retain only coarser structure, such as the identity of the occupied Lorenz scroll. The third is generality: whether the observed behaviour is a property of one well-studied system or reproduces across a heterogeneous collection of chaotic dynamics.

\section{Related work}
\label{sec:related-works}

Classical chaotic forecasting combines state-space reconstruction with local approximations of the underlying dynamics. Delay-coordinate reconstruction provides a route from scalar observations to an equivalent state-space representation~\cite{Takens1981}, while local analog and nonlinear models use nearby reconstructed states to predict subsequent evolution \cite{FarmerSidorowich1987,Casdagli1989}. Neural dynamic-map approximation and broader nonlinear-time-series analyses developed the same forecasting perspective in related forms \cite{PrincipeRathieKuo1992,KantzSchreiber2004}. Multistep forecasts are commonly assembled recursively, fitted separately for individual horizons, or produced as joint multiple-output vectors~\cite{benTaieb2012review}. Researchers have also studied these strategies with neural predictors for chaotic time series~\cite{SangiorgioDercole2020}. They differ in error propagation and output structure, but they do not treat a configuration of deliberately omitted intermediate positions as part of a recursively generated prediction.

Most directly related is the line of partial multistep prediction based on non-successive observations. Gromov and Borisenko introduced predictable and non-predictable observations and used patterns of non-successive positions in recursive multistep prediction \cite{GromovBorisenko}. Gromov and Baranov subsequently formalized partial multistep prediction as a two-objective problem balancing error on predictable positions against their non-predictable fraction, and introduced possible prediction trajectories together with a pointwise oracle construction \cite{GromovBaranov}. A later self-healing method iteratively revisits unresolved positions with the aim of restoring additional values \cite{GromovTomashchukRukavishnikov2024}. These studies establish the partial-prediction framework and particular trajectory constructions. Here we instead ask how the admissible family of partial trajectories is distributed and organized across masks and lead times.

Modern full-output forecasting of chaotic dynamics provides a complementary benchmark context. Representative approaches include recurrent neural networks~\cite{vlachas2018lstm}, reservoir computing \cite{pathak2018modelFree}, and explicit nonlinear lagged-feature maps such as next-generation reservoir computing~\cite{gauthier2021ngrc}. Reported prediction horizons cannot, however, be compared independently of the sampling protocol. Reservoir and random-feature studies show that predictive performance changes with temporal resolution and that the preferred cadence depends on the predictor construction \cite{tsuchiyama2023temporalResolution,mandal2025hitrun}. Section~\ref{sec:experimental-setup} specifies the observation intervals and Lyapunov-time normalization used here.

Recent studies have scaled this benchmark perspective to large collections of chaotic systems. Time-series foundation models produce competitive zero-shot forecasts of chaotic systems unseen during training~\cite{ZhangGilpin2025}, and models pretrained purely on synthetic chaotic dynamics extend this direction~\cite{LaiBaoGilpin2025}. Gilpin reviews generative machine-learning approaches to nonlinear dynamics~\cite{Gilpin2024}. A systematic comparison across chaotic benchmarks also found that carefully tuned local and lightweight methods remain competitive with deep architectures~\cite{Schotz2026}. The pattern-based local predictors used here belong to this lightweight class, and the object of study is the structure of partial trajectories built on top of such a predictor rather than the predictor itself.

Selective prediction provides a related language for withholding unreliable outputs. Selective classification and regression with a reject option trade conditional risk against coverage \cite{GeifmanElYaniv2017,DenisHebiriZaoui2020}. Recent bounded-abstention work extends this idea to structured, correlated outputs in multi-horizon time-series forecasting~\cite{StradiottiEtAl2026}. We consider a different information structure. Omitting an intermediate position changes the contexts used to generate subsequent recursive predictions, so irregular active sets produce distinct partial trajectories rather than different abstention decisions over a jointly produced multi-horizon output.

The Lorenz analysis also draws on established spatial and symbolic descriptions of the attractor. The classical treatment of the Lorenz equations describes its two-scroll geometry~\cite{Sparrow1982}, while kneading and symbolic dynamics describe the topological organization of Lorenz-type chaos~\cite{BarrioShilnikovShilnikov2012}. These references motivate scroll identity as a coarse descriptor of the trajectory.

For evaluation beyond a single canonical system, the \dysts{} collection provides an interpretable set of heterogeneous chaotic dynamics for common-protocol forecasting comparisons~\cite{Gilpin2021Dysts}. It therefore supplies a suitable setting for testing whether an effect observed on the Lorenz series persists for complete multivariate states across systems.

The closest partial-prediction studies do not characterize the error distribution over randomly sampled admissible masks, and they do not track the complete threshold-swept error--non-predictability trade-off as a function of lead time. We address these gaps through random-mask sampling and an agreement-based acceptance rule, and we use the block-wise oracle search to probe whether a long-horizon low-error region exists at all. We examine two remaining questions, the topological content of the surviving predictions and the transfer of the construction across systems, within the same mask-conditioned framework.

\section{Problem statement}
\label{sec:problem-statement}

We consider a normalized scalar chaotic time series $Y=\{y_t\}_{t\ge 0}$, $y_t\in\mathbb R$, after removal of the transient part of the underlying dynamics. Let $t$ be the last observed position, let $H\in\mathbb N$ be the maximal prediction horizon, and let $I_{t,H}=\{t+1,\ldots,t+H\}$ be the corresponding future interval. Let $h\in\{1,\ldots,H\}$ be the discrete lead time, so that the target position is $t+h$.

Following \citet{GromovBaranov}, the prediction algorithm associates with each target position a finite multiset of candidate values,
\begin{equation}
\widehat S_{t+h} = \left\{ \widehat y_{t+h}^{(1)}, \ldots, \widehat y_{t+h}^{(N_{t+h})} \right\}.
\end{equation}
The multiset may contain repeated numerical values. We retain candidates produced by different patterns or partial trajectories as separate contributions.

For an algorithmic configuration $\theta\in\Theta$, predictability is represented by
\begin{equation}
\zeta(\widehat S_{t+h};\theta) =
\begin{cases}
1, & \text{predictable},\\
0, & \text{non-predictable}.
\end{cases}
\end{equation}
We set $\zeta(\varnothing;\theta)=0$. If $\zeta(\widehat S_{t+h};\theta)=1$, the unified prediction is
\begin{equation}
\widehat y_{t+h}(\theta) = g(\widehat S_{t+h};\theta).
\end{equation}
The rule may reject a prediction because no candidate is available or because the available candidates do not support a sufficiently reliable unified value.

Let $\chi\in\{\mathrm{val},\mathrm{test}\}$, and let $\mathcal D_{\chi,h}$ be the set of starting positions evaluated at lead time $h$ in segment $\chi$. The corresponding predictable and non-predictable sets are
\begin{equation}
\mathcal P_{\chi,h}(\theta) =
\left\{ t\in\mathcal D_{\chi,h}: \zeta(\widehat S_{t+h};\theta)=1 \right\},
\end{equation}
\begin{equation}
\mathcal N_{\chi,h}(\theta) =
\mathcal D_{\chi,h}\setminus\mathcal P_{\chi,h}(\theta).
\end{equation}
The first objective is the non-predictable fraction
\begin{equation}
I_1^\chi(h,\theta) = \frac{|\mathcal N_{\chi,h}(\theta)|} {|\mathcal D_{\chi,h}|}.
\end{equation}

The second objective measures prediction error conditionally on $\mathcal P_{\chi,h}(\theta)$. For $\mathcal P_{\chi,h}(\theta)\neq\varnothing$, define
\begin{equation}\label{eq:i2rmse}
I_{2,\mathrm{RMSE}}^\chi(h,\theta) =
\left[ \frac{1}{|\mathcal P_{\chi,h}(\theta)|} \smashoperator[r]{\sum_{t\in\mathcal P_{\chi,h}(\theta)}} \bigl(\widehat y_{t+h}(\theta)-y_{t+h}\bigr)^2 \right]^{1/2}.
\end{equation}
With
\begin{equation}
\bar y_{\chi,h}(\theta) = \frac{1}{|\mathcal P_{\chi,h}(\theta)|} \sum_{t\in\mathcal P_{\chi,h}(\theta)}y_{t+h},
\end{equation}
the $R^2$-based criterion is
\begin{equation}\label{eq:i2r2}
I_{2,R^2}^\chi(h,\theta) = 1-R_{\chi,h}^2(\theta) =
\frac{ \displaystyle \sum_{t\in\mathcal P_{\chi,h}(\theta)} \bigl(\widehat y_{t+h}(\theta)-y_{t+h}\bigr)^2 }
{ \displaystyle \sum_{t\in\mathcal P_{\chi,h}(\theta)} \bigl(y_{t+h}-\bar y_{\chi,h}(\theta)\bigr)^2 }.
\end{equation}
We omit configurations for which a conditional metric is undefined from the corresponding comparison.

For either metric $m\in\{\mathrm{RMSE},R^2\}$, partial multistep prediction is the componentwise minimization problem
\begin{equation}
\min_{\theta\in\Theta} \mathbf I_m^\chi(h,\theta), \qquad \mathbf I_m^\chi(h,\theta) =
\left( I_1^\chi(h,\theta), I_{2,m}^\chi(h,\theta) \right).
\end{equation}
The transformation $I_{2,R^2}=1-R^2$ serves only to cast the $R^2$-based problem into the same minimization orientation. In this formulation, $I_1^\chi$ is the cost of selectivity. Sweeping algorithmic configurations traces retained accuracy against the fraction of positions given up, and the resulting Pareto fronts quantify this trade-off as a function of lead time.

We distinguish two mechanisms: trajectory-level skipping and ensemble-level rejection. An active set of retained future positions defines a single partial trajectory, and the construction skips the complementary positions (Sec.~\ref{subsec:trajectory-masks}). This trajectory-level mask is distinct from $\mathcal N_{\chi,h}(\theta)$, which contains the starting positions that the final ensemble rule rejects at a fixed lead time.

\section{Methods}
\label{sec:methods}

\subsection{Pattern-based local prediction}
\label{subsec:pattern-local-prediction}

Let $L\ge2$ be the generalized $z$-vector length and let
\begin{equation}
\alpha = \left( k_1^{(\alpha)},\ldots,k_{L-1}^{(\alpha)} \right)
\in \aleph(L,K_{\max}) = \{1,\ldots,K_{\max}\}^{L-1}.
\end{equation}
The generalized $z$-vector starting at position $m$ is
\begin{equation}
z_m^{(\alpha)} = \left( y_m, y_{m+k_1^{(\alpha)}}, \ldots, y_{m+\sum_{j=1}^{L-1}k_j^{(\alpha)}} \right),
\end{equation}
with the consecutive vector corresponding to $\alpha=(1,\ldots,1)$.

For each pattern, the first $L-1$ coordinates form a context $x_m^{(\alpha)}$ and the final coordinate is its target $u_m^{(\alpha)}$.
\begin{equation}
\begin{aligned}
x_m^{(\alpha)} &= \left( y_m,\; y_{m+k_1^{(\alpha)}},\; \ldots,\; y_{m+\sum_{j=1}^{L-2} k_j^{(\alpha)}} \right),\\
u_m^{(\alpha)} &= y_{m+\sum_{j=1}^{L-1} k_j^{(\alpha)}},\\
\end{aligned}
\end{equation}
Thus, $z_m^{(\alpha)} = \left(x_m^{(\alpha)}, u_m^{(\alpha)} \right).$

Let $\mathcal T_{\mathrm{train}}$ be the training index set specified by the experimental split, and let $\mathcal I_\alpha$ contain all starting positions for which the complete vector lies in that segment. We use these context--target pairs to fit a pattern-specific candidate operator
\begin{equation}
\mathcal M_\alpha: \mathbb R^{L-1} \longrightarrow \{\text{finite multisets in }\mathbb R\}.
\end{equation}

To predict the value at position $\tau$, we align the pattern so that its target position coincides with $\tau$. The corresponding context is
\begin{equation}
\widetilde x_\tau^{(\alpha)} = x_m^{(\alpha)},
\qquad m=\tau-\sum_{j=1}^{L-1}k_j^{(\alpha)},
\end{equation}
assembled from the observed history and, during a rollout, from previously predicted active positions. The operator returns
\begin{equation}
\widehat S_\tau^{(\alpha)} =
\mathcal M_\alpha\!\left(\widetilde x_\tau^{(\alpha)}\right),
\end{equation}
and we pool the available pattern-specific candidates by multiset union:
\begin{equation}
\widehat S_\tau = \biguplus_{\alpha\in\aleph(L,K_{\max})} \widehat S_\tau^{(\alpha)}.
\end{equation}
The local unified value is their arithmetic mean,
\begin{equation}
g_{\mathrm{loc}}(S) = \frac{1}{|S|} \sum_{v\in S}v.
\end{equation}

We compare three local-predictor classes. The closest-motifs operator returns the targets of training contexts within a prescribed Euclidean neighbourhood, with a nearest-context fallback when that neighbourhood is empty. A decision tree or random forest defines a deterministic scalar map $f_\alpha$, embedded in the multiset representation as $\mathcal M_\alpha(x)=\{f_\alpha(x)\}$. Closest motifs may contribute several occurrences for one pattern, whereas a tree or forest contributes a singleton. Appendix~\ref{app:predictor-selection} reports exact predictor specifications and the parameter-selection grids.

\subsection{Trajectory-level masks and masked prediction}
\label{subsec:trajectory-masks}

A nonempty \emph{active set} $A\subseteq I_{t,H}$ specifies a partial trajectory, and $\Sigma_A=I_{t,H}\setminus A$ is its \emph{skipped complement}. The block-mask sampler generates active sets according to $A\sim\mathcal S_{m_{\max}}(t,H)$, which partitions the future interval into random consecutive blocks and retains one randomly selected position from each block. Appendix~\ref{app:block-mask-sampler} gives the complete sampler, its density and gap properties, and its boundary treatment. Write $\mathcal A_{m_{\max}}(t,H)$ for the class of nonempty active sets obtainable by this construction.

For a pattern $\alpha$, let
\begin{equation}
C_\tau^{(\alpha)} = \left\{ \tau-\sum_{j=r}^{L-1}k_j^{(\alpha)}: r=1,\ldots,L-1 \right\}
\end{equation}
be the positions used in its context at $\tau$. At an active position, the pattern is usable only if none of these positions is skipped. The candidate multiset is
\begin{equation}
\widehat S_{\tau,A} = \biguplus_{\substack{ \alpha\in\aleph(L,K_{\max})\\
C_\tau^{(\alpha)}\cap\Sigma_A=\varnothing }} \mathcal M_\alpha\!\left(\widetilde x_\tau^{(\alpha)}\right).
\end{equation}

Given the observed history $\eta_t=(y_0,\ldots,y_t)$, the fixed family $\{\mathcal M_\alpha\}$ determines the unified partial trajectory $\widehat\xi_A$ pointwise:
\begin{equation}
\widehat\xi_A(\tau) = g_{\mathrm{loc}}(\widehat S_{\tau,A}), \qquad \tau\in A.
\end{equation}
The rollout processes active positions in increasing order, and their predicted values may enter later contexts. It assigns no value at positions in $\Sigma_A$. We write
\begin{equation}
\widehat\xi_A = \mathcal R(A;\eta_t)
\end{equation}
for the map returning this trajectory.

\subsection{Oracle block-wise mask search}
\label{subsec:oracle-daemons}

For a nonempty active set, the oracle active-position error of a trajectory $\widehat\xi$ against the observed series $y$ is
\begin{equation}
\mathcal E_A(\widehat\xi) = \left[ \frac{1}{|A|} \sum_{\tau\in A} \bigl(\widehat\xi(\tau)-y_\tau\bigr)^2 \right]^{1/2}.
\end{equation}
The formal optimum in the admissible mask class is
\begin{equation}
A^\ast \in \operatorname*{arg\,min}_{A\in\mathcal A_{m_{\max}}(t,H)} \mathcal E_A\!\left(\mathcal R(A;\eta_t)\right).
\end{equation}
This optimum defines the existence question but is not computed exhaustively.

The block-wise daemon\footnote{In the internal slang of this line of research, oracle constructions of this kind are called \emph{daemons}: Socrates is known to have had his own daemon, who gave him advice in his most painful situations. Below we refer to the construction as the oracle search.} performs a finite stochastic search. From the current partial history, it samples admissible masks within a residual window, selects the candidate with the smallest window-level oracle RMSE, appends the selected active positions and predictions, and continues over the remaining horizon. The search repeats the complete construction for several trials and returns the pair with the smallest full-horizon active-position RMSE:
\begin{equation}
(\widehat A,\widehat\xi) \in \operatorname*{arg\,min}_{1\le q\le N_{\mathrm{tr}}} \mathcal E_{A^{(q)}}\!\left(\widehat\xi^{(q)}\right).
\end{equation}
Thus, $\widehat A$ is the best mask found by the finite search, not a guaranteed realization of $A^\ast$. Every returned active set belongs to $\mathcal A_{m_{\max}}(t,H)$. Appendix~\ref{app:block-daemon} gives the search parameters, pseudocode, and the admissibility argument. A comparison with the pointwise oracle construction of \citet{GromovBaranov} appears in Appendix~\ref{app:pointwise-daemon}.

The search evaluates masks using the true continuation and serves only as an oracle existence instrument, not as a deployable prediction procedure.

\subsection{Statistical characterization of random partial trajectories}
\label{subsec:random-partial-trajectory-distribution}

Let $\mathcal D_{\mathrm{dist}}$ be the set of starting positions used in the random-mask experiment. For every $t\in\mathcal D_{\mathrm{dist}}$ and $k=1,\ldots,N_{\mathrm{dist}}$, independently sample
\begin{equation}
A_t^{(k)}\sim\mathcal S_{m_{\max}}(t,H)
\end{equation}
and construct
\begin{equation}
\widehat\xi_t^{(k)} = \mathcal R(A_t^{(k)};\eta_t).
\end{equation}
Neither mask sampling nor rollout construction uses the true continuation.

For $h\le H$, let $A_{t,h}^{(k)}=A_t^{(k)}\cap I_{t,h}$. Whenever this prefix is nonempty, its error is
\begin{equation}
R_{t,h}^{(k)} = \mathcal E_{A_{t,h}^{(k)}} \left( \widehat\xi_t^{(k)}\big|_{A_{t,h}^{(k)}} \right).
\end{equation}
For $p\in(0,1)$, quantiles summarize the mask-wise error distribution separately at each starting position:
\begin{equation}
q_{p,t,h} = Q_p\!\left(\{R_{t,h}^{(k)}\}_k\right),
\end{equation}
and we then summarize these quantiles across starting positions by
\begin{equation}
\widetilde q_{p,h} = \operatorname{median}_{t\in\mathcal D_{\mathrm{dist}}}q_{p,t,h},
\end{equation}
where we omit empty prefixes.

A second diagnostic measures disagreement among partial trajectories that are active at the evaluated position. Let
\begin{equation}
\mathcal V_{t,h} = \left\{ \widehat\xi_t^{(k)}(t+h): t+h\in A_t^{(k)} \right\},
\end{equation}
and set
\begin{equation}
\nu_{t,h} = \operatorname{std}(\mathcal V_{t,h}).
\end{equation}
Across starting positions, this spread is summarized as
\begin{equation}
\widetilde\nu_h = \operatorname{median}_{t\in\mathcal D_{\mathrm{dist}}}\nu_{t,h}.
\end{equation}

\subsection{Ensemble of partial trajectories}
\label{subsec:ensemble-partial-trajectories}

At each starting position, sample
\begin{equation}
A_t^{(k)} \sim \mathcal S_{m_{\max}}(t,H),
\qquad k=1,\ldots,\Nens.
\end{equation}
Independent masks generate the trajectory pool. At lead time $h$, every active trajectory contributes its complete pattern-level candidate multiset:
\begin{equation}
\widehat\Xi_{t,h}^{\mathrm{part}} = \biguplus_{\substack{ 1\le k\le \Nens\\
t+h\in A_t^{(k)} }} \widehat S_{t+h,A_t^{(k)}}.
\end{equation}
We retain all candidate occurrences and their multiplicities.

\subsection{Perturbed trajectory generator}
\label{subsec:perturbed-generator}

The perturbed trajectory generator produces a family of ordinary recursive rollouts. It predicts every intermediate future position, keeps it active, and diversifies recursive histories through scaled additive noise. The construction follows the randomly perturbed trajectories of \citet{GromovBaranov}. It provides a reference mechanism for the partial generator. Both produce a pool of candidate values at each future position, but here the histories differ through noise rather than through omission.

For the $k$-th perturbed trajectory, suppose that a recursive history is available up to a position $\tau-1$. The fixed family of local candidate operators produces noiseless pattern-level predictions at $\tau$. We retain these candidates as the trajectory's contribution to the ensemble candidate pool. We use a noiseless per-trajectory mean $\widehat y_\tau^{(k)}$, computed from the pattern-level predictions, only to form the value written into the recursive history.

We inject randomness after computing this noiseless mean. The value written into the history is
\begin{equation}
\widetilde y_\tau^{(k)} = \widehat y_\tau^{(k)}
+ \drel\sigma_{\mathrm{train}}\xi_\tau^{(k)}, \qquad \xi_\tau^{(k)}\sim\mathcal N(0,1),
\end{equation}
where $\sigma_{\mathrm{train}}$ is the training standard deviation of the normalized signal, $\drel\in[0,1]$ is the relative perturbation scale, and $\xi_\tau^{(k)}$ are independent across positions and trajectories. The coefficient $\drel$ is a dimensionless normalizing factor that sets the noise amplitude as a fraction of $\sigma_{\mathrm{train}}$, not a standard deviation. Each recursive step applies the perturbation once, and it affects only the value stored in the history, so the candidates passed to the ensemble aggregation remain noiseless.

For a starting position $t$ and lead time $h$, the pooled candidate multiset $\widehat\Xi_{t,h}^{\mathrm{pert}}$ is the multiset union of the per-trajectory contributions over the sampled perturbed trajectories.

With $\drel=0$, all rollouts coincide and the construction reduces to one deterministic recursive rollout. This \emph{recursive baseline} predicts at every position without rejection and is the reference method in the predictor-selection and cross-system comparisons.

\subsection{Aggregation and acceptance rule}
\label{subsec:aggregation-acceptance}

Either generator returns, at a starting position $t$ and lead time $h$, a pooled candidate multiset, written $\widehat\Xi_{t,h}$ with the generator index dropped when it is clear from context. For $\widehat\Xi_{t,h}\neq\varnothing$, the pooled center and mean absolute deviation are
\begin{equation}
c_{t,h} = \frac{1}{|\widehat\Xi_{t,h}|} \sum_{v\in\widehat\Xi_{t,h}}v,
\qquad d_{t,h} =
\frac{1}{|\widehat\Xi_{t,h}|} \sum_{v\in\widehat\Xi_{t,h}}|v-c_{t,h}|.
\end{equation}
The ensemble predictability rule, the \emph{agreement rule} below, is
\begin{equation}\label{eq:agreement}
\zeta_{\mathrm{ens}} (\widehat\Xi_{t,h};\teps) =
\begin{cases}
1, & d_{t,h}<\teps \ \text{and}\ \widehat\Xi_{t,h}\neq\varnothing,\\
0, & \text{otherwise}.
\end{cases}
\end{equation}
For an accepted prediction,
\begin{equation}
\widehat y_{t+h} = g_{\mathrm{ens}}(\widehat\Xi_{t,h}) = c_{t,h}.
\end{equation}
Along a given error--non-predictability curve, only the dispersion threshold $\teps$ varies. The local predictor, trajectory generator, aggregation rule, and ensemble size remain fixed.

For a vector-valued system $\mathbf y_t\in\mathbb R^D$, each pattern-specific predictor returns a complete vector, and a trajectory contributes either all components at $t+h$ or none. We compute centers and absolute deviations componentwise, with point-level dispersion
\begin{equation}
d_{t,h}^{\mathrm{point}} = \frac{1}{D} \sum_{j=1}^{D}d_{t,h,j}.
\end{equation}
The rule accepts the complete vector when $d_{t,h}^{\mathrm{point}}<\teps$.

In the notation of Sec.~\ref{sec:problem-statement},
\begin{equation}
\begin{aligned}
\widehat S_{t+h}&\equiv\widehat\Xi_{t,h}, &\theta&=(\bar\theta,\teps),\\
\zeta&=\zeta_{\mathrm{ens}}, &g&=g_{\mathrm{ens}}.
\end{aligned}
\end{equation}
Here $\bar\theta$ contains the fixed local predictor, the trajectory generator with its settings (the mask sampler or the perturbation scale), the ensemble size, and the aggregation rule.

\paragraph{Oracle ensemble.}
We construct the oracle ensemble from matched starting positions, sampled masks, candidate multisets, and dispersion threshold. It differs only in the weights assigned to partial trajectories. For trajectory $k$, the oracle score is its full-horizon active-position RMSE,
\begin{equation}
E_{t,k}^{\mathrm{orc}} = \mathcal E_{A_t^{(k)}}\!\left(\widehat\xi_t^{(k)}\right).
\end{equation}
If $r_t(k)$ is the rank of this error and we assign rank one to the smallest value, the trajectory weight is
\begin{equation}
w_{t,k}^{\mathrm{orc}} = \frac{ \exp\!\left[-\gorc(r_t(k)-1)\right] }
{ \displaystyle \sum_{j=1}^{\Nens} \exp\!\left[-\gorc(r_t(j)-1)\right] }.
\end{equation}
Each candidate occurrence inherits the weight of its parent trajectory. Appendix~\ref{app:oracle-ensemble-aggregation} gives the weighted center, weighted dispersion, and lead-time normalization.

\subsection{Evaluation metrics and Pareto fronts}
\label{subsec:evaluation-metrics}

For a fixed threshold and lead time, write $\mathcal P=\mathcal P_{\chi,h}(\teps)$. We evaluate all conditional metrics on this set. The conditional root-mean-square error $\operatorname{RMSE}_{\chi,h}$ and the conditional coefficient of determination $R^2_{\chi,h}$ are, respectively, $I_{2,\mathrm{RMSE}}^\chi$ and $1-I_{2,R^2}^\chi$ of Eqs.~\eqref{eq:i2rmse} and~\eqref{eq:i2r2}, evaluated under the correspondence of Sec.~\ref{subsec:aggregation-acceptance}. We suppress threshold dependence when no ambiguity arises.

Let $\widetilde\Theta$ be the threshold grid specified by the experimental protocol. Sweeping $\teps\in\widetilde\Theta$ produces an RMSE-based trade-off between $I_1^\chi$ and conditional RMSE and an $R^2$-based trade-off between $I_1^\chi$ and conditional $R^2$. We write $\mathcal F_{\chi,h}^{\mathrm{RMSE}}$ and $\mathcal F_{\chi,h}^{R^2}$ for their nondominated subsets. The first minimizes both coordinates. The second minimizes $I_1^\chi$ and maximizes $R^2$. The construction applies unchanged to either trajectory generator and to the oracle-ensemble reduction.

For the normalized Lorenz series, let $\mu$ be the fixed scroll-separation level and define
\begin{equation}
s(u) =
\begin{cases}
+1, & u\ge\mu,\\
-1, & u<\mu.
\end{cases}
\end{equation}
The correctly classified predictable starting positions are
\begin{equation}
\mathcal G_{\chi,h} = \left\{ t\in\mathcal P: s(\widehat y_{t+h})=s(y_{t+h}) \right\}.
\end{equation}
Scroll accuracy is
\begin{equation}
\operatorname{ScrollAcc}_{\chi,h} =
\frac{|\mathcal G_{\chi,h}|}{|\mathcal P|},
\end{equation}
and the error conditional on correct scroll identification is
\begin{equation}
\operatorname{RMSE}^{\mathrm{CS}}_{\chi,h} =
\left[ \frac{1}{|\mathcal G_{\chi,h}|} \sum_{t\in\mathcal G_{\chi,h}} (\widehat y_{t+h}-y_{t+h})^2 \right]^{1/2}.
\end{equation}

For a vector-valued system, we compute $R^2$ separately for each state component and summarize it by
\begin{equation}\label{eq:r2min}
R^2_{\chi,h,\min} = \min_{1\le j\le D}R^2_{\chi,h,j}.
\end{equation}
For a scalar series this reduces to $R^2_{\chi,h}$. Let $H_{\mathrm{eval}}$ be the largest evaluated discrete lead time. Given a maximum admissible non-predictable fraction $\rho_{\max}$, the ensemble positive-$R^2$ horizon at threshold $\teps$ is
\begin{equation}\label{eq:hplus}
\begin{aligned}
&H_{\chi}^{+,\mathrm{ens}} (\teps;\rho_{\max}) = \max\Bigl\{ q\in\{0,\ldots,H_{\mathrm{eval}}\}: \\
&\qquad R^2_{\chi,h,\min} (\teps)>0, \quad I_1^{\chi} (h,\teps)\le\rho_{\max}, \\
&\qquad \text{for every }h=1,\ldots,q \Bigr\}.
\end{aligned}
\end{equation}
If the condition holds over the complete evaluated range, the horizon equals $H_{\mathrm{eval}}$. The baseline horizon $H_\chi^{+,\mathrm{base}}$ follows the same positive-prefix rule for the recursive baseline, which predicts at every intermediate position. For the cross-system comparison, the relative extension is
\begin{equation}\label{eq:qh}
Q_H = \frac{H_{\mathrm{test}}^{+,\mathrm{ens}}} {H_{\mathrm{test}}^{+,\mathrm{base}}},
\end{equation}
whenever the baseline horizon is positive.

\subsection{Threshold selection}
\label{subsec:threshold-selection}

We select thresholds on the validation segment by one of two rules.

The horizon-first rule maximizes the validation horizon of Sec.~\ref{subsec:evaluation-metrics}. Set
\begin{equation}
H_{\max}^+(\rho_{\max}) = \max_{\teps\in\widetilde\Theta} H_{\mathrm{val}}^{+,\mathrm{ens}} (\teps;\rho_{\max})
\end{equation}
and
\begin{equation}
\mathcal C(\rho_{\max}) = \operatorname*{arg\,max}_{\teps\in\widetilde\Theta} H_{\mathrm{val}}^{+,\mathrm{ens}} (\teps;\rho_{\max}).
\end{equation}
For $H_{\max}^{+}(\rho_{\max})>0$, among the thresholds attaining this horizon, the principal threshold minimizes the mean validation non-predictable fraction over the attained prefix:
\begin{equation}\label{eq:eps-H}
\teps_H^\ast \in \operatorname*{arg\,min}_{\teps\in\mathcal C(\rho_{\max})} \frac{1}{H_{\max}^+(\rho_{\max})} \sum_{h=1}^{H_{\max}^+(\rho_{\max})} I_1^{\mathrm{val}}(h,\teps).
\end{equation}

The selective rule targets a designated lead time $h_{\mathrm{sel}}$:
\begin{equation}\label{eq:eps-sel}
\teps_{\mathrm{sel}}^\ast \in \operatorname*{arg\,max}_{\substack{ \teps\in\widetilde\Theta\\
I_1^{\mathrm{val}}(h_{\mathrm{sel}},\teps) \le\rho_{\max} }} R^2_{\mathrm{val},h_{\mathrm{sel}}} (\teps).
\end{equation}
We hold a selected threshold fixed over lead time. The experimental setup specifies the values of $\rho_{\max}$ and $h_{\mathrm{sel}}$ and the threshold grid.

\section{Experimental setup}
\label{sec:experimental-setup}

\subsection{Data generation, preprocessing, and splits}
\label{subsec:data-setup}

The scalar experiments use the $x$-coordinate of the Lorenz system
\begin{equation}
\dot x=\sigma(y-x), \qquad \dot y=x(\rho-z)-y,
\qquad \dot z=xy-\beta z,
\end{equation}
with
\begin{equation}
\sigma=10, \qquad \rho=28,
\qquad \beta=\frac{8}{3}.
\end{equation}
We integrate the system with the fourth-order Runge--Kutta method and use the observation interval $\Delta t_{\mathrm{obs}}=0.1$. We discard the first 10\,000 observations, globally min--max normalize the remaining scalar series, and then extract the experimental segments.

The training, validation, and test segments contain 10\,000, 20\,000, and 30\,000 observations, respectively. The train--validation gap is 20\,000 observations, and the validation--test gap is 10\,000 observations. We use the training segment to fit the local predictors and the validation segment to select predictors and perform the structural Lorenz experiments. The cross-system study evaluates a deployment-style transfer of validation-selected thresholds to unseen test data.

For the Lorenz series, we use the estimate~\cite{Sprott2003}
\begin{equation}
\lambda_{\max}\approx0.91, \qquad \Lambda_h =
\lambda_{\max}\Delta t_{\mathrm{obs}}h = 0.091h.
\end{equation}
We report all numerical lead times and prediction horizons in the prose, tables, and figure captions in Lyapunov units, and retain the discrete indices $h$ and $H$ in algorithmic definitions.

The cross-system experiment uses the complete state vectors of the non-delay continuous-time systems in the fixed \dysts{} version 0.96 snapshot used for the experiments \cite{Gilpin2021Dysts}. Delay-differential systems are excluded upstream by the snapshot's own classification. The snapshot lists 129 such systems. One of them (\emph{HyperLu}) failed integration under the protocol below, so we omit it and evaluate 128 systems. For each system, we discard the first 100 Lyapunov times and set the observation interval to
\begin{equation}
\Delta t_{\mathrm{obs}} = \frac{0.1}{\lambda_{\max}},
\end{equation}
so that one observation corresponds to 0.1 Lyapunov times. We globally min--max normalize each state component separately. Each system uses the same segment lengths and gaps as the Lorenz series. We place starting positions uniformly over the admissible positions of the segment used in each experiment.

\subsection{Shared predictor and mask configuration}
\label{subsec:shared-configuration}

We select the local-predictor configuration on the Lorenz validation segment. We compare the pattern parameters using one-step and multistep baseline-rollout RMSE and prefer a compact pattern family that retains adequate temporal reach. The selected configuration is $L=5$, $K_{\max}=5$. At this configuration, we compare closest motifs, decision trees, and random forests under the same validation protocol. The one-step (0.091-Lyapunov-time) validation RMSE is 0.0046 for the decision tree, 0.0058 for the random forest, and 0.0276 for closest motifs with $\varepsilon_{\mathrm{motif}}=0.02$. The tree and forest remain close over the short-horizon baseline regime, with the tree generally giving the smaller rollout RMSE, and all three recursive rollouts converge to a common high-error regime at large lead times (Fig.~\ref{fig:local-predictor-comparison}, Appendix~\ref{app:predictor-selection}). We retain an unrestricted decision tree with minimum leaf size one for the structural and cross-system experiments. The random forest is ten unrestricted trees with the same minimum leaf size. Appendix~\ref{app:predictor-selection} reports the exact predictor definitions and complete parameter grids.

The block-mask sampler generates partial trajectories with $m_{\max}=3$. For scalar series, each pattern-specific tree returns one scalar candidate. In the cross-system experiment, we use the same construction in multi-output form. Each context contains complete $D$-dimensional states, and the tree returns a vector in $\mathbb R^D$. We fix $L$, $K_{\max}$, and $m_{\max}$ across all systems.

\subsection{Structural analysis on the Lorenz series}
\label{subsec:lorenz-structural-setup}

Table~\ref{tab:lorenz-structural-setup} summarizes the configurations of the block-wise oracle search, random-mask distribution study, and ensemble and oracle-ensemble experiments.

\begin{table*}[tbp]
\centering
\caption{Configuration of the structural studies on the Lorenz validation segment. Numerical lead times are reported in Lyapunov units.}
\label{tab:lorenz-structural-setup}
\begin{tabular}{lccc}
\toprule
Study
&
Starting positions
&
Maximum lead time
&
Sampling or search size
\\
\midrule
Block-wise oracle search
&
100
&
27.3
&
multiple search configurations
\\
Random-mask distributions
&
100
&
13.7
&
$N_{\mathrm{dist}}=40\,000$
\\
Ensemble and oracle ensemble
&
5\,000
&
13.7
&
$\Nens=100$
\\
\bottomrule
\end{tabular}
\end{table*}

The oracle-search and distribution studies use the same set of 100 starting positions. We run the oracle search over a fixed finite collection of combinations of $H_{\mathrm{blk}}$, $N_{\mathrm{cand}}$, and $N_{\mathrm{tr}}$, as described in Appendix~\ref{app:block-daemon}. At each starting position, the search returns the trial with the smallest full-horizon active-position RMSE. We classify a search result as successful when $\mathcal E_{\widehat A}(\widehat\xi)<0.01$.

For each starting position in the distribution study, we generate one pool of $N_{\mathrm{dist}}=40\,000$ partial trajectories up to 13.7 Lyapunov times and reuse it for all discrete prefixes.

We evaluate the ensemble and oracle ensemble at the same 5\,000 starting positions using matched mask pools. We sweep the ensemble threshold over
\begin{equation}
\widetilde\Theta = \{ 0, 10^{-4}, 2\cdot10^{-4}, \ldots, 0.3 \}.
\end{equation}
At every lead time, we extract Pareto fronts from all defined threshold outcomes without an additional coverage prefilter. The oracle ensemble uses $\gorc=0.04$, which we hold fixed in the main structural comparison. Appendix~\ref{app:oracle-weight-sensitivity} reports additional rank concentrations.

We select the principal and selective thresholds by the rules of Eqs.~\eqref{eq:eps-H} and~\eqref{eq:eps-sel} with
\begin{equation}
\rho_{\max}=0.90, \qquad h_{\mathrm{sel}}=60,
\qquad \Lambda_{h_{\mathrm{sel}}}=5.5.
\end{equation}
We apply each selected threshold without re-estimation to the oracle-ensemble reduction of the same mask pool.

The comparison between the partial and perturbed generators uses the same 5\,000 starting positions and $\Nens=100$ trajectories per generator. We evaluate the perturbed generator over the relative perturbation scales
\begin{equation}
\drel \in \{0.001, 0.005, 0.010, 0.020, 0.035, 0.080\},
\end{equation}
and report the comparison at
\begin{equation}
h\in\{33,55,77\}, \qquad \Lambda_h\approx\{3.0,5.0,7.0\}.
\end{equation}
For analyses at fixed operating points, we apply the threshold-selection rules of Sec.~\ref{subsec:threshold-selection} to each generator separately. Dispersion scales differ between the generators, so we do not compare raw threshold values across generators. In the scroll diagnostics, the scroll-separation level $\mu$ is the mean of the globally normalized scalar Lorenz series.

\subsection{Cross-system evaluation protocol}
\label{subsec:dysts-setup}

The cross-system study uses the vector-valued ensemble, not the oracle ensemble. We retain the shared predictor and mask configuration of Sec.~\ref{subsec:shared-configuration}. For each system, the ensemble contains $\Nens=50$ partial trajectories, which we evaluate over windows of up to 80 Lyapunov times.

We evaluate the recursive baseline and the ensemble at 1\,000 uniformly placed starting positions on each validation and test segment. The two methods use matched starting positions within each split. We calculate prediction quality for each state component and summarize it by the minimum componentwise $R^2$ of Eq.~\eqref{eq:r2min}.

For each system, we select the ensemble threshold on the validation segment by the horizon-first rule of Eq.~\eqref{eq:eps-H}, with $\rho_{\max}=0.95$. We apply the selected threshold to the test segment without re-estimation. The reported quantities are
\begin{equation}
H_{\mathrm{test}}^{+,\mathrm{base}},
\qquad H_{\mathrm{test}}^{+,\mathrm{ens}}, \qquad Q_H,
\end{equation}
with both horizons expressed in Lyapunov units.

Several systems in the collection are quasi-periodic or only weakly chaotic under the present integration protocol. For such systems, the recursive-baseline positive-$R^2$ horizon reaches 40 Lyapunov times. The evaluated window does not resolve the horizon of predictability, and the comparison carries no information about prediction beyond it. We exclude systems whose baseline horizon reaches this threshold. Two further systems (\emph{ArnoldWeb} and \emph{PanXuZhou}) have a recursive baseline whose $R^2$ is nonpositive at the first prediction step, so they have no defined baseline horizon, and we exclude them as well. Together this removes 21 of the 128 evaluated systems, leaving 107 retained for cross-system analysis. Appendix~\ref{app:excluded-systems} lists the excluded systems, together with a representative trajectory of one of them.

\section{Results}
\label{sec:results}

\subsection{Long low-error partial trajectories}
\label{subsec:results-daemon}

\begin{figure}[t]
    \centering
    \includegraphics[width=\columnwidth]
    {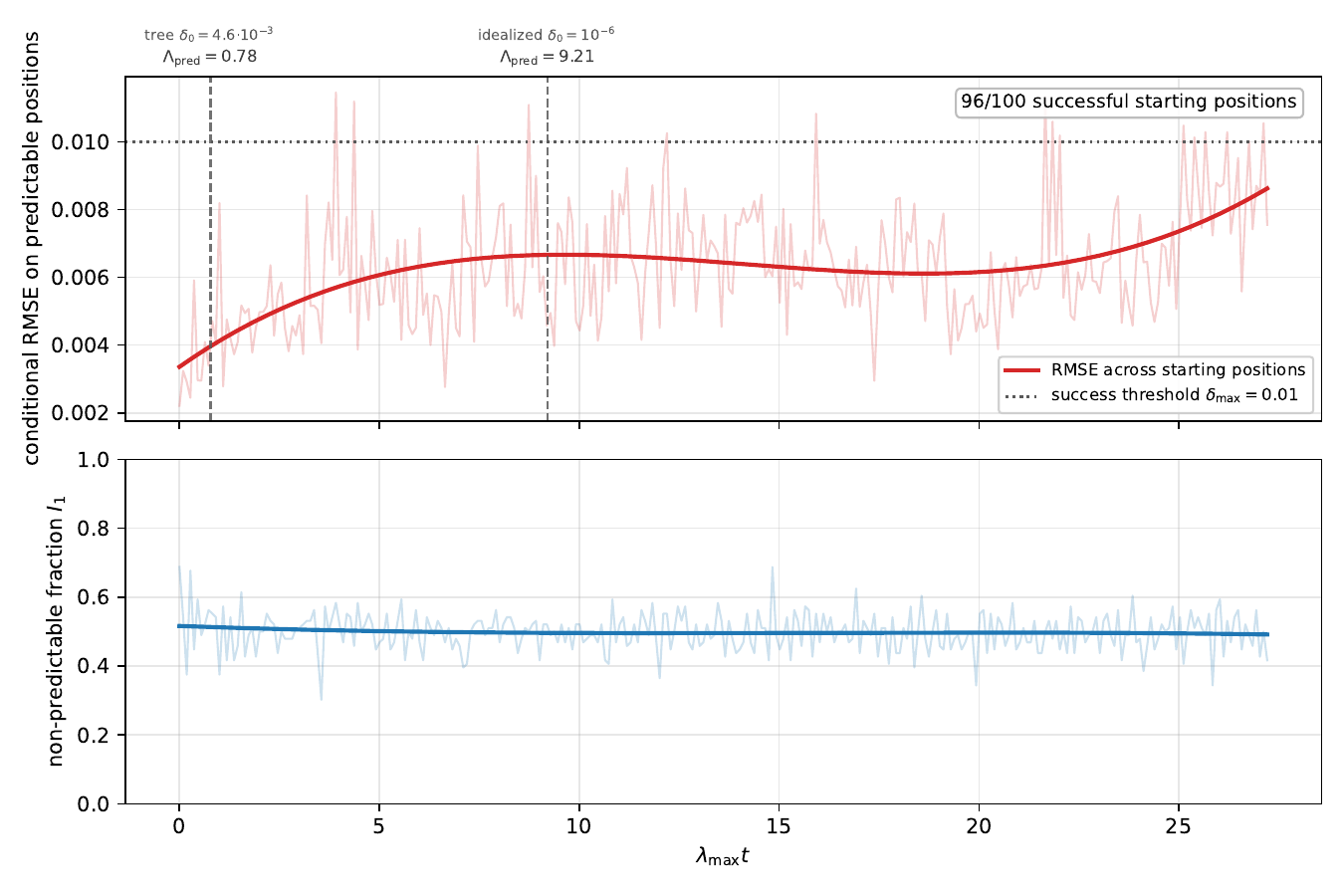}
    \caption{Block-wise oracle-search results for the 96 successful Lorenz starting positions over 27.3 Lyapunov times. The upper panel shows RMSE across successful starting positions for which the evaluated position is active. The thin line is the raw statistic; the thick red curve is a smoothing spline through it. The lower panel shows the fraction of successful partial trajectories in which each lead time is skipped. The dashed vertical lines mark the horizon estimate of Eq.~\eqref{eq:tpred} for the success level $\delta_{\max}=0.01$, placed at the one-step validation RMSE of the tree predictor ($\delta_0=0.0046$, 0.8 Lyapunov times) and at an idealized one-step error more than three orders of magnitude smaller ($\delta_0=10^{-6}$, 9.2 Lyapunov times).}
    \label{fig:daemon-plateau}
\end{figure}

The oracle search finds a partial trajectory with full-horizon active-position RMSE below 0.01 for 96 of the 100 examined starting positions over 27.3 Lyapunov times. At each lead time, we evaluate RMSE over the successful starting positions for which that position is active. The lower panel reports the complementary skipped fraction (Fig.~\ref{fig:daemon-plateau}).

\begin{figure*}[tbp]
    \centering
    \includegraphics[width=0.78\textwidth]
    {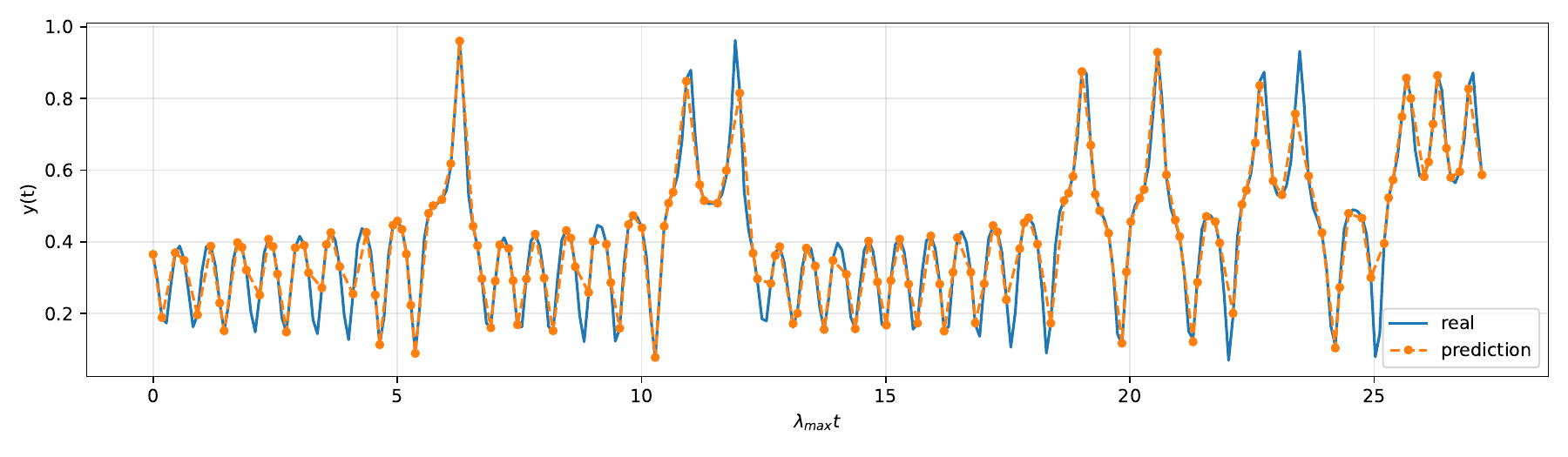}
    \caption{Representative partial trajectory returned by the block-wise oracle search. The solid curve is the globally normalized Lorenz series and the markers are predictions at the active positions of the selected mask. The finite oracle search selects the mask using active-position RMSE over the complete prediction interval.}
    \label{fig:daemon-example}
\end{figure*}

The skipped fraction fluctuates around one half, as expected from the block-mask sampler, and does not increase systematically with lead time. The low error does not result from a progressive disappearance of active positions. Figure~\ref{fig:daemon-example} shows a representative returned trajectory. The predictions at the active positions follow the series over the complete interval, while the skipped positions remain spread across it. The result is an existence statement, obtained under a finite search budget and with access to the true continuation.

\subsection{Error distribution of random partial trajectories}
\label{subsec:results-random-distribution}

\begin{figure*}[t]
    \centering
    \includegraphics[width=1\textwidth]
    {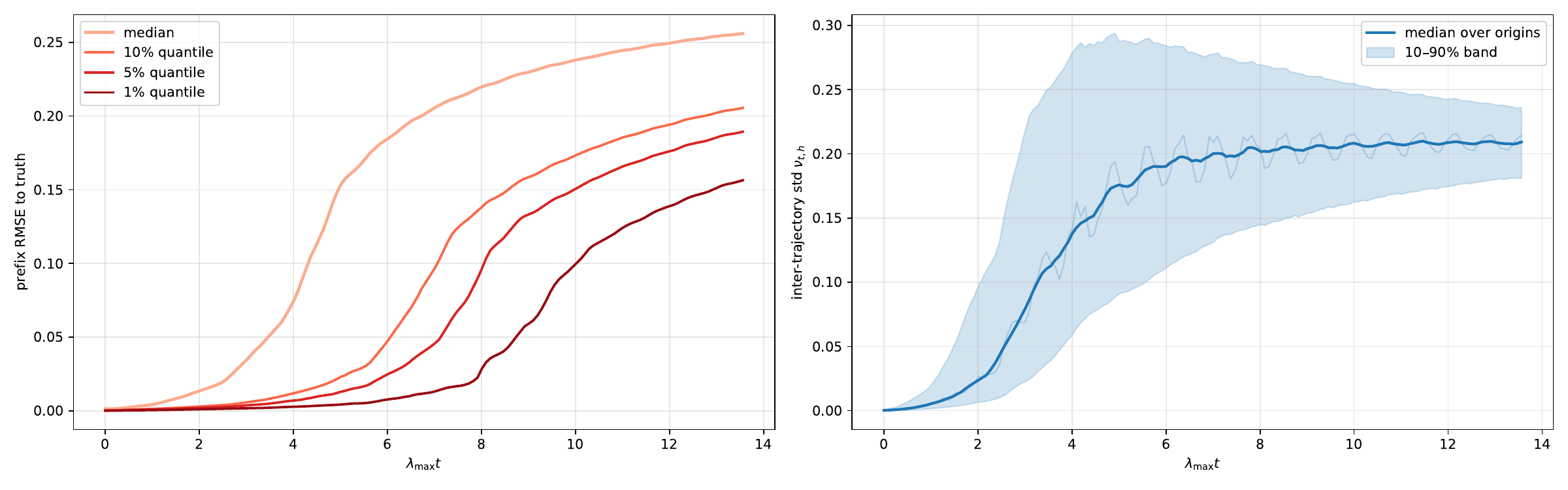}
    \caption{Error distribution of randomly sampled partial trajectories on the Lorenz validation segment. For each of 100 starting positions, the experiment samples $N_{\mathrm{dist}}=40\,000$ masks once up to 13.7 Lyapunov times and reuses them for every discrete prefix. Panel (a) shows the median across starting positions of the 1\%, 5\%, 10\%, and 50\% prefix-RMSE quantiles, computed separately at each starting position. Panel (b) shows the median across starting positions of the inter-trajectory standard deviation $\nu_{t,h}$ and its 10--90\%
    range; the thin line is the raw median and the thick line its smoothed trend. Horizontal axes give lead time in Lyapunov units.}
    \label{fig:random-mask-quantiles}
\end{figure*}

The oracle search shows that favourable masks exist, but not how common they are among random masks. Figure~\ref{fig:random-mask-quantiles}(a) compares the median prefix RMSE with the 1\%, 5\%, and 10\% lower quantiles computed separately at each starting position. The median error begins to rise after the short-horizon regime and approaches a broad long-horizon plateau. The lower quantiles increase more gradually and remain well separated from the median. Random masks induce a heterogeneous error distribution in which the typical partial trajectory loses accuracy before the lower empirical tail.

The inter-trajectory standard deviation grows over the same intermediate lead-time range (Fig.~\ref{fig:random-mask-quantiles}(b)). After the initial concentrated regime, the sampled partial trajectories spread over increasingly different continuations, and their dispersion remains high at longer lead times.

Together, a low-error tail and broadly separated continuations motivate aggregation by agreement. Uniform averaging cannot distinguish mutually concentrated candidate values from incompatible regions of the pool, whereas the ensemble dispersion supplies a pointwise criterion for retaining only sufficiently concentrated predictions.

\subsection{Evolution of the ensemble Pareto front}
\label{subsec:results-pareto}

\begin{figure}[!t]
    \centering
    \includegraphics[width=\columnwidth]
    {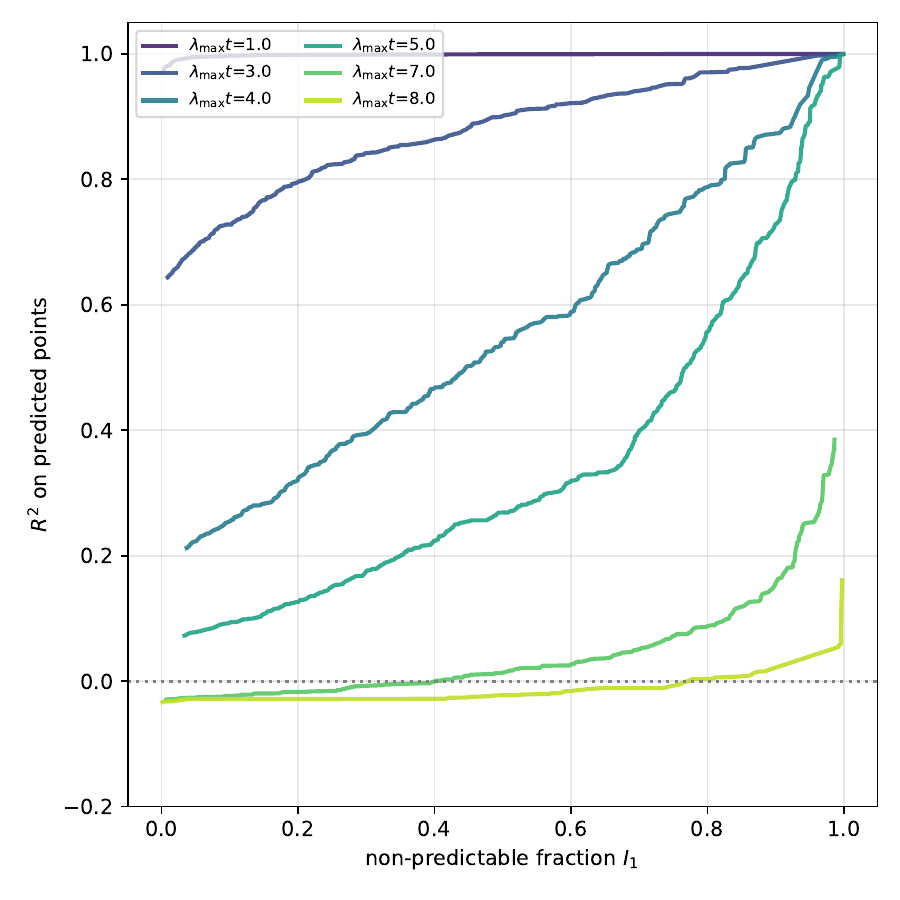}
    \caption{$R^2$--non-predictability trade-offs of the ensemble on the Lorenz validation segment. The ensemble contains $\Nens=100$ partial trajectories and is evaluated at 5\,000 starting positions. Each curve contains only nondominated points from the complete threshold grid. Colours identify lead times in Lyapunov units.}
    \label{fig:ensemble-pareto-r2}
\end{figure}

Sweeping the agreement threshold gives a nondominated trade-off between conditional $R^2$ and the non-predictable fraction, shown in Fig.~\ref{fig:ensemble-pareto-r2}. At short lead times, the fronts rise steeply and then saturate after rejecting only a limited fraction of starting positions. Geometric prediction is available for a broad portion of the validation segment, and rejection mainly removes a relatively small set of difficult starts.

As lead time approaches the range between 4 and 5 Lyapunov times, the effect of increasing selectivity changes. Improvements in conditional $R^2$ become distributed over a much wider range of $I_1$, and at later lead times only strongly selected subsets retain positive geometric skill. Predominantly convex long-horizon fronts replace the predominantly concave short-horizon fronts. We refer to this change in Pareto-front geometry as the transition at $\Lambda^\ast$.

At the principal operating point $\teps_H^\ast=0.138$, the ensemble retains positive $R^2$ up to 7.55 Lyapunov times. The oracle-ensemble reduction of the same mask pools, with the same threshold, retains positive $R^2$ throughout the entire 13.7-Lyapunov-time interval. The sampled pool therefore carries predictive information over an interval almost twice as long as the interval recovered by the ensemble, and the gap between the two horizons measures what uniform aggregation leaves unextracted.

Appendix~\ref{app:ensemble-results} reports the RMSE-based and oracle-ensemble fronts. They provide error-scale and oracle reference views of the same threshold sweep, while we use the ensemble $R^2$ front for the main structural analysis.

\subsection{Partial versus perturbed ensembles}
\label{subsec:results-partial-perturbed}

The perturbed generator tests whether the Lorenz effect comes from masking specifically, or from aggregating diversified recursive histories more generally. We compare the two generators under the acceptance rule of Eq.~\eqref{eq:agreement}, using matched starting positions and ensemble sizes. We evaluate the perturbed family over the perturbation-scale grid of Sec.~\ref{subsec:lorenz-structural-setup}.

\begin{figure}[tbp]
  \centering
  \includegraphics[width=\columnwidth]
  {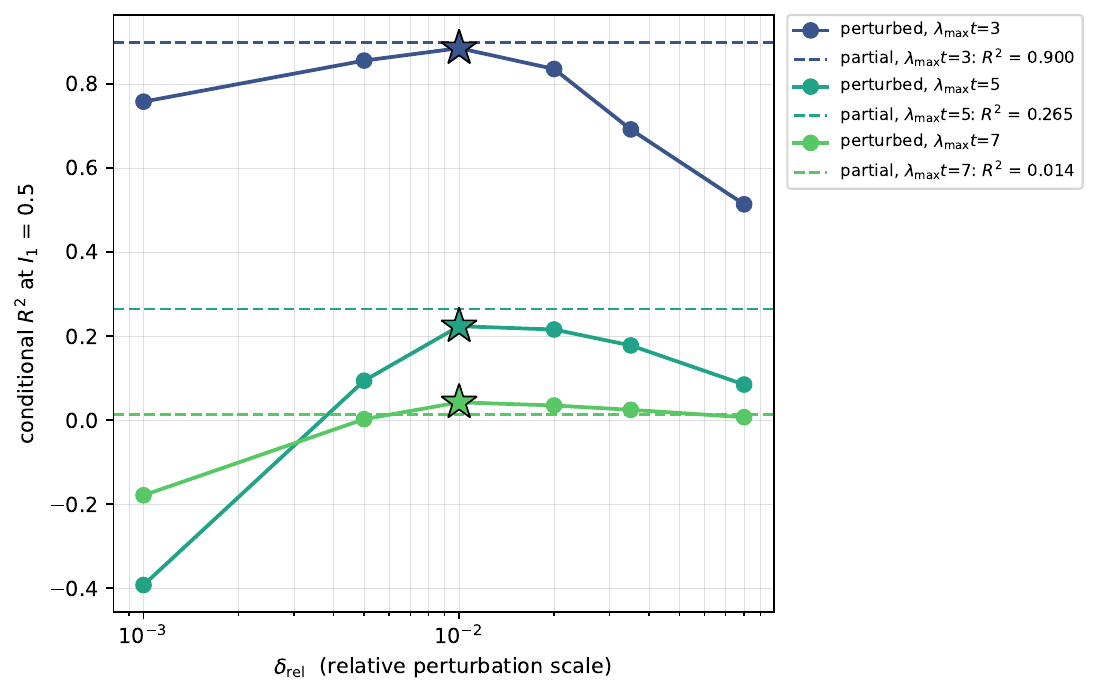}
  \caption{Sensitivity of the perturbed generator to the relative perturbation scale. Curves report conditional $R^2$ at $I_1=0.5$ as $\drel$ is varied, with the partial trajectory ensemble shown as the corresponding reference level at each displayed lead time. Stars mark the best perturbation scale at each lead time.}
  \label{fig:perturbed-sigma-curve}
\end{figure}

At a fixed non-predictable fraction $I_1=0.5$, the partial ensemble is not uniformly better across the displayed lead times (Fig.~\ref{fig:perturbed-sigma-curve}). At $\Lambda_h\approx3.0$, the best perturbed value and the partial value are almost tied: $R^2=0.885$ and $R^2=0.900$, respectively. At $\Lambda_h\approx5.0$, the partial ensemble gives $R^2=0.265$ against $R^2=0.224$ for the tuned perturbed generator, and the comparison is not limited to this coverage level. The complete $R^2$--non-predictability front of the partial ensemble lies above every perturbed front at this lead time (Appendix~\ref{app:partial-perturbed-fronts}). At $\Lambda_h\approx7.0$, both generators are close to the loss of positive predictive skill, with $R^2=0.014$ for the partial ensemble and $R^2=0.042$ for the best perturbed value. The partial ensemble concentrates its advantage at intermediate lead times and achieves it without any noise tuning.

The perturbed family is controlled by its noise scale. Across the three displayed lead times, the scale that produces the best perturbed curve is $\drel=0.010$ (Fig.~\ref{fig:perturbed-sigma-curve}), and the dependence on the scale is not monotonic. Tuning matters most at intermediate lead times. At $\Lambda_h\approx5.0$, $\drel=0.001$ gives $R^2=-0.392$, the tuned value $\drel=0.010$ gives $R^2=0.224$, and $\drel=0.080$ lowers the same metric to $R^2=0.085$. The perturbed generator is useful only in a narrow part of the perturbation grid.

\subsection{From geometric to topological predictability}
\label{subsec:results-topological-transition}

For the scalar Lorenz series, the scroll label defined in Sec.~\ref{subsec:evaluation-metrics} provides an operational two-region partition motivated by the spatial and symbolic descriptions of Lorenz dynamics~\cite{Sparrow1982,BarrioShilnikovShilnikov2012}. This subsection examines what the accepted predictions of both generators retain after pointwise accuracy degrades.

We evaluate each generator at the two operating points selected by the threshold-selection rules of Sec.~\ref{subsec:threshold-selection} applied to that generator. For the partial ensemble, the principal and selective thresholds are $\teps_H^\ast=0.138$ and $\teps_{\mathrm{sel}}^\ast=0.065$. The perturbed generator uses $\drel=0.010$ with its own selected thresholds.

\begin{figure*}[t]
    \centering
    \includegraphics[width=0.98\textwidth]
    {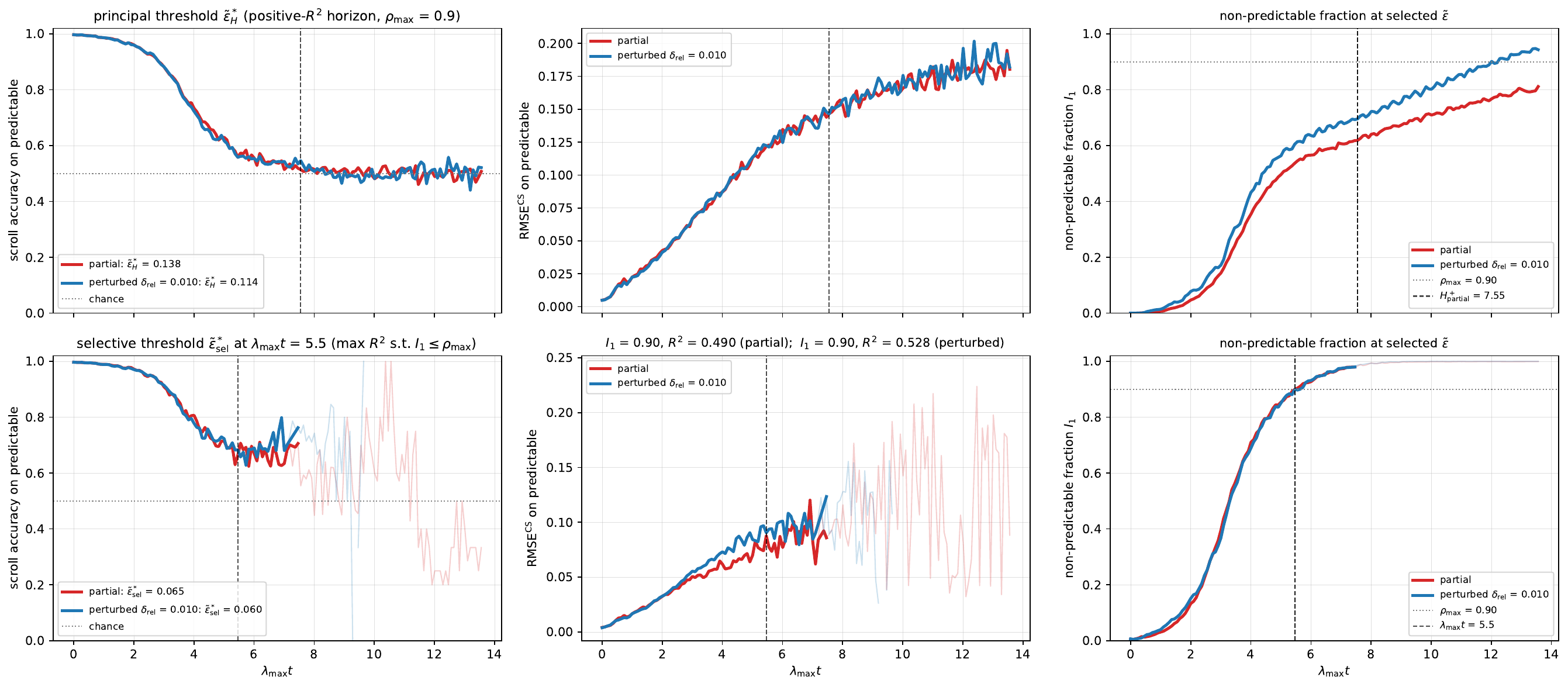}
    \caption{Scroll identity and within-scroll localization at selected operating points. The upper row uses the principal threshold, which maximizes the positive-$R^2$ horizon with $\rho_{\max}=0.9$. The lower row uses the selective threshold chosen at $h_{\mathrm{sel}}=60$ ($\Lambda_{h_{\mathrm{sel}}}\approx5.5$ Lyapunov times). In each row, the partial trajectory ensemble and the perturbed generator with $\drel=0.010$ use thresholds selected by the same rule. Left panels show scroll accuracy on accepted predictions, middle panels show correct-scroll RMSE, and right panels show the non-predictable fraction. Accepted predictions retain scroll identity after within-scroll localization error has begun to increase.}
    \label{fig:protocol-scroll-rmse}
\end{figure*}

Both operating points give the same picture (Fig.~\ref{fig:protocol-scroll-rmse}). As lead time grows past $\Lambda^\ast$, correct-scroll RMSE begins to rise, but scroll accuracy does not collapse with it. The accepted prediction is then no longer a reliable point estimate of the future state, yet it still identifies the occupied scroll well above chance over the retained range. For the partial ensemble, the error conditional on correct scroll identification is approximately 0.14 at its positive-$R^2$ horizon of 7.55 Lyapunov times under the principal threshold, and approximately 0.09 at 5.5 Lyapunov times under the selective threshold. The oracle-ensemble scroll diagnostics at the same operating points and representative individual predictions appear in Appendices~\ref{app:oracle-scroll} and~\ref{app:lorenz-examples}.

Both generators share this behaviour. For the scroll partition used here, the predictions lose geometrically precise state information while retaining topological information about the occupied scroll.

\subsection{Generalization across chaotic systems}
\label{subsec:results-cross-system}

\begin{figure}[t]
    \centering
    \includegraphics[width=\columnwidth]
    {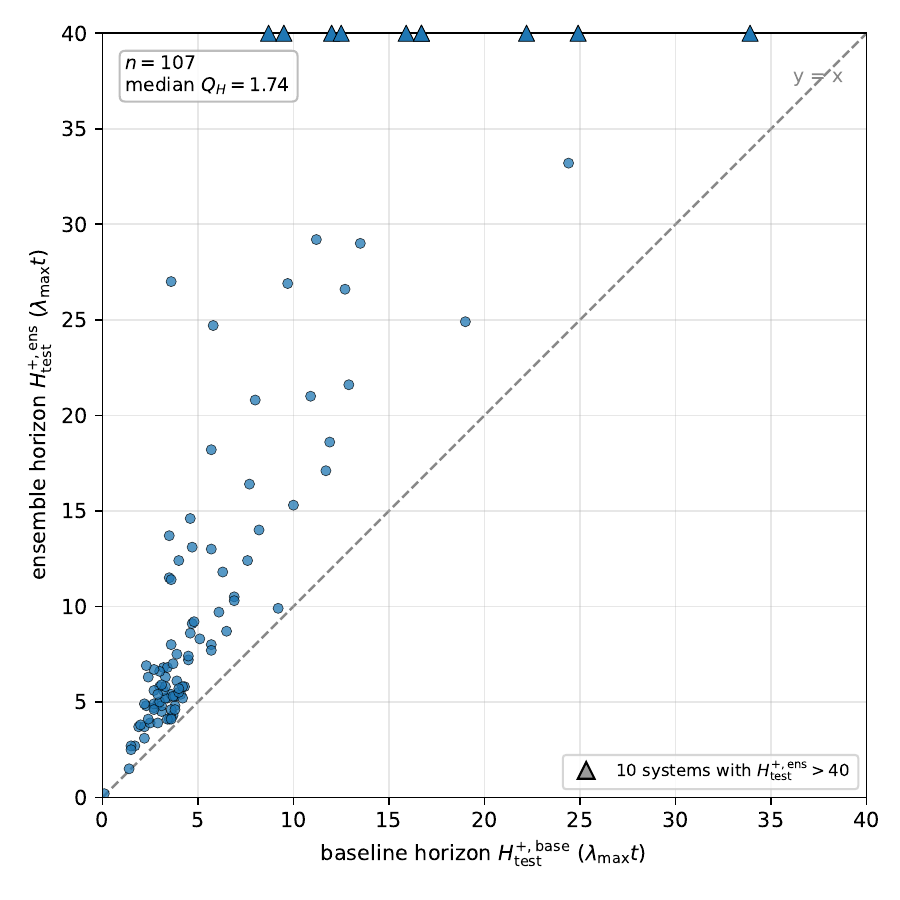}
    \caption{Recursive-baseline and ensemble positive-$R^2$ horizons for the complete state vectors of the 107 retained chaotic systems. Each point represents one system. We compute $R^2$ separately for every state component and summarize it by the minimum. For each system, we select the threshold on 1\,000 validation starting positions under $\rho_{\max}=0.95$ and apply it without re-estimation to the matched 1\,000 test starting positions. The ensemble contains $\Nens=50$ partial trajectories. The dashed diagonal marks equal horizons. Triangles at the upper edge mark the 10 systems whose ensemble horizon exceeds 40 Lyapunov times. Both axes give Lyapunov times.}
    \label{fig:cross-system-horizons}
\end{figure}

We compare the recursive baseline and the partial-trajectory ensemble for the complete state vectors of the 107 systems of the \dysts{} collection~\cite{Gilpin2021Dysts} retained by the protocol of Sec.~\ref{subsec:dysts-setup}. Figure~\ref{fig:cross-system-horizons} reports the comparison. We evaluate each system on the test segment with its validation-selected threshold.

The ensemble strictly extends the positive-$R^2$ horizon for every retained system:
\begin{equation}
H_{\mathrm{test}}^{+,\mathrm{ens}} > H_{\mathrm{test}}^{+,\mathrm{base}}.
\end{equation}
The median horizon ratio $Q_H$ [Eq.~\eqref{eq:qh}] across the retained systems is 1.74. The magnitude of the extension is system dependent. Some systems remain close to the diagonal, whereas others gain several Lyapunov times.

\section{Conclusions}
\label{sec:conclusion}

Beyond the horizon of predictability the recursive rollout fails, but the series does not become uniformly unpredictable. Accurate partial continuations exist far past that point, part of them recoverable without future observations, and what survives changes in kind, from geometric accuracy to topological structure, rather than vanishing. This is what partial multistep prediction gains, at the cost of conceding a growing fraction of positions as non-predictable~\cite{GromovBaranov}. Our construction recovers it from the agreement of candidates sampled over different active sets.

The block-wise oracle search finds accurate continuations out to some 27 Lyapunov times, far beyond the recursive-baseline range, but candidate agreement recovers only part of them without access to the future. The oracle reduction of the same pools stays predictive over a substantially longer range, although this reference is optimistic because the oracle weights use the true continuation. Weighting trajectories by an observable quality is the most direct route to narrowing this gap.

The transition at $\Lambda^\ast$, between 4 and 5 Lyapunov times, separates the two regimes: up to it the Pareto fronts are concave, and a broad accepted subset keeps geometric accuracy at moderate rejection. Beyond it they turn convex: positive geometric skill requires increasingly strong selection, and what survives is topological. Accepted predictions still identify the occupied Lorenz scroll after within-scroll localization has degraded, so the continuous geometry of the attractor is lost before its symbolic identity. Both masking and perturbation show this survival under the same acceptance rule, so it is a property of that rule. The partial ensemble reaches it without the tuned noise scale the perturbed control requires.

For the retained systems of the \dysts{} collection, under one predictor, one mask sampler, and one ensemble configuration, the agreement ensemble strictly extends the positive-$R^2$ horizon of the recursive baseline, with a median horizon ratio of 1.74 and a system-dependent magnitude. A local linear predictor and the closest-motifs operator reproduce the same concave-to-convex change, and their accepted predictions also retain positive conditional $R^2$ at long lead times (Appendix~\ref{app:predictor-robustness}). The transition is robust across predictors: only the location of $\Lambda^\ast$ shifts with the choice of local predictor.

These results extend the operational horizon of the recursive baseline by trading coverage for conditional accuracy. The construction needs only a mask-conditioned operator that returns candidate values at selected positions. Natural next steps are observable trajectory weighting to close the extraction gap, mask-aware predictors that remove the dependence on pattern enumeration, and pretrained forecast models as candidate generators.

\begin{acknowledgments}
This research was supported in part through computational resources of HPC facilities at HSE University. Support from the Basic Research Program of HSE University is gratefully acknowledged (HSE-BR-2025-025).
\end{acknowledgments}

\section*{Conflict of Interest}

The authors declare that they have no conflicts of interest.

\clearpage

\appendix

\section{Algorithmic details}
\label{app:algorithms}

\subsection{Block-mask sampler}
\label{app:block-mask-sampler}

Algorithm~\ref{alg:block-mask-sampler} gives the sampler used throughout the partial-trajectory experiments. The sampler partitions the future interval into consecutive blocks with independently sampled lengths, selects one active position uniformly from each block, and discards any position beyond the right boundary. It repeats the complete procedure only when it produces an empty active set.

\begin{algorithm}[H]
\caption{Block-mask sampler $\mathcal S_{m_{\max}}(t,H)$}
\label{alg:block-mask-sampler}
\begin{algorithmic}[1]
\Require Starting position $t$, horizon $H$, maximum block length $m_{\max}$
\Ensure Nonempty active set $A\subseteq I_{t,H}$

\Repeat
    \State $A\gets\varnothing$
    \State $p\gets t+1$
    \While{$p\le t+H$}
        \State Draw $\ell\sim\operatorname{Uniform}\{1,\ldots,m_{\max}\}$
        \State Draw $u\sim\operatorname{Uniform}\{0,\ldots,\ell-1\}$
        \State $\tau\gets p+u$
        \If{$\tau\le t+H$}
            \State $A\gets A\cup\{\tau\}$
        \EndIf
        \State $p\gets p+\ell$
    \EndWhile
\Until{$A\neq\varnothing$}

\State \Return $A$
\end{algorithmic}
\end{algorithm}

Away from the right boundary, each block contributes one active position. The asymptotic expected active fraction is
\begin{equation}
\frac{\mathbb E|A|}{H} \approx \frac{1}{\mathbb E\ell} =
\frac{2}{m_{\max}+1}.
\end{equation}
For $m_{\max}=3$, approximately one half of the future positions are active. If $a_r<a_{r+1}$ are selected in consecutive blocks, then
\begin{equation}
1\le a_{r+1}-a_r\le 2m_{\max}-1.
\end{equation}
Thus, the sampler produces irregular masks while imposing a finite upper bound on the distance between successive active positions. With the configuration used in the experiments,
\begin{equation}
2m_{\max}-1=K_{\max}=5,
\end{equation}
so every realized gap between consecutive active positions is representable by the generalized-pattern family. The final, possibly truncated block affects only the right-boundary correction and not the preceding gap bound.

\subsection{Block-wise stochastic search}
\label{app:block-daemon}

The block-wise search uses the true continuation only to compare sampled masks. Let $\eta\oplus\widehat\xi$ be the extension of a partial history $\eta$ by the newly retained predictions of $\widehat\xi$. In every residual window, the algorithm samples $N_{\mathrm{cand}}$ active sets, keeps the candidate with the smallest window-level active-position RMSE, and appends its predictions to the current partial history. The algorithm repeats the complete construction $N_{\mathrm{tr}}$ times.

\begin{algorithm}[H]
\caption{Block-wise stochastic oracle search}
\label{alg:block-wise-daemon}
\begin{algorithmic}[1]
\Require Starting position $t$, observed history $\eta_t$, true continuation
$y$, horizon $H$, residual-window length $H_{\mathrm{blk}}$, sampler
$\mathcal S_{m_{\max}}$, $N_{\mathrm{cand}}$, and $N_{\mathrm{tr}}$
\Ensure Active set $\widehat A$, partial trajectory $\widehat\xi$, and
full-horizon active-position error $\widehat E$

\For{$q=1,\ldots,N_{\mathrm{tr}}$}
    \State $\eta^{(q)}\gets\eta_t$
    \State $A^{(q)}\gets\varnothing$
    \State $\widehat\xi^{(q)}\gets\varnothing$
    \State $j\gets t$

    \While{$j<t+H$}
        \State $\ell\gets\min\{H_{\mathrm{blk}},\,t+H-j\}$
        \For{$r=1,\ldots,N_{\mathrm{cand}}$}
            \State $A_r\sim\mathcal S_{m_{\max}}(j,\ell)$
            \State $\widehat\xi_r\gets\mathcal R(A_r;\eta^{(q)})$
            \State $e_r\gets\mathcal E_{A_r}(\widehat\xi_r)$
        \EndFor
        \State $r^\ast\gets
        \operatorname*{arg\,min}_{1\le r\le N_{\mathrm{cand}}}e_r$
        \State $A^{(q)}\gets A^{(q)}\cup A_{r^\ast}$
        \State $\widehat\xi^{(q)}\gets
        \widehat\xi^{(q)}\oplus\widehat\xi_{r^\ast}$
        \State $\eta^{(q)}\gets\eta^{(q)}\oplus\widehat\xi_{r^\ast}$
        \State $j\gets\max A_{r^\ast}$
    \EndWhile

    \State $E_q\gets\mathcal E_{A^{(q)}}(\widehat\xi^{(q)})$
\EndFor

\State $\widehat q\gets
\operatorname*{arg\,min}_{1\le q\le N_{\mathrm{tr}}}E_q$
\State $\widehat A\gets A^{(\widehat q)}$
\State $\widehat\xi\gets\widehat\xi^{(\widehat q)}$
\State $\widehat E\gets E_{\widehat q}$
\State \Return $\widehat A,\widehat\xi,\widehat E$
\end{algorithmic}
\end{algorithm}

The search evaluates a finite collection of configurations
\begin{equation}
\mathcal C_{\mathrm{search}} =
\left\{ (H_{\mathrm{blk}}^{(c)}, N_{\mathrm{cand}}^{(c)}, N_{\mathrm{tr}}^{(c)}) \right\}_{c=1}^{C}.
\end{equation}
At each starting position, the search runs all configurations and retains the trajectory with the smallest full-horizon active-position RMSE.

The search evaluates the local criterion only on positions active in the current residual window and selects the final trial by active-position RMSE over the complete prediction interval. These two objectives play distinct roles. The first determines the next block, whereas the second selects among complete trials.

Every set $A^{(q)}$ returned by a trial belongs to $\mathcal A_{m_{\max}}(t,H)$. The block-mask sampler generates each accepted subset on the current residual interval. Its realized block sequence can be truncated after the block containing its last active position. Concatenating the truncated sequences from successive residual windows gives a valid block construction on $I_{t,H}$ with active set exactly $A^{(q)}$. Consequently,
\begin{equation}
A^{(q)}\in\mathcal A_{m_{\max}}(t,H),
\qquad \widehat A\in\mathcal A_{m_{\max}}(t,H).
\end{equation}
Thus, $\widehat A$ is the best admissible active set found by the finite search, not a guaranteed realization of the formal optimum $A^\ast$.

Figure~\ref{fig:daemon-example} shows a representative result. The selected predictions remain close to the true series over the active positions, while skipped positions prevent rejected intermediate values from entering later contexts.

\subsection{Pointwise and block-wise oracle searches}
\label{app:pointwise-daemon}

The pointwise oracle search follows \citet{GromovBaranov}. Starting from $A_0=\varnothing$, it constructs a provisional prediction at $t+i$ from the observed history and the predictions retained at positions in $A_{i-1}$. The search accepts a provisional value $\widehat y_{t+i}=g_{\mathrm{loc}}(\widehat S_{t+i\mid A_{i-1}})$ when its candidate multiset is nonempty and $|\widehat y_{t+i}-y_{t+i}|<\varepsilon_{\mathrm{orc}}$. The search then updates the active set by
\begin{equation}
A_i=
\begin{cases}
A_{i-1}\cup\{t+i\}, & \text{if accepted},\\
A_{i-1}, & \text{otherwise}.
\end{cases}
\end{equation}
Rejected values do not enter later contexts, and the true continuation is never inserted into the prediction history. The resulting active set is $A_{\mathrm{GB}}=A_H$.

This rule is greedy and local. The search accepts or rejects each position using its own provisional error. The block-wise search instead compares complete masks inside a residual window and selects a mask by their joint active-position error. Both procedures return the same type of trajectory-level object, but they search for it by different mechanisms.

\begin{figure*}[tbp]
\centering
\includegraphics[width=0.96\textwidth]
{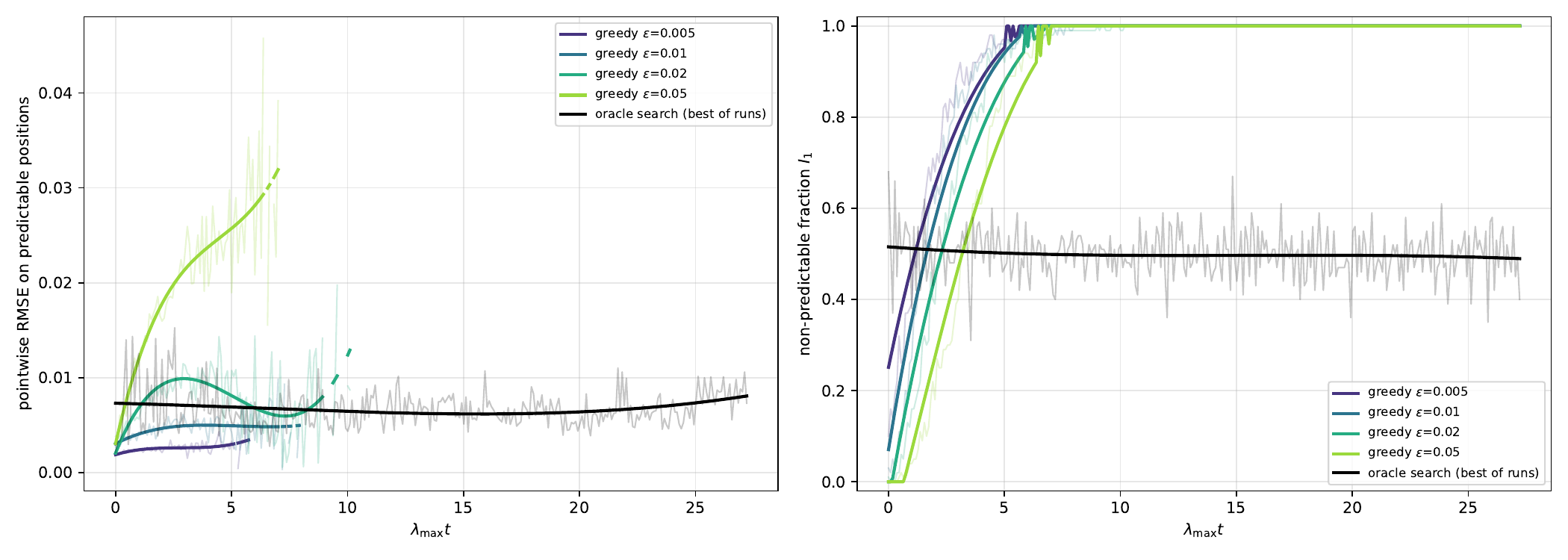}
\caption{Comparison of the pointwise oracle search (labelled \emph{greedy}) with the block-wise search on the Lorenz validation segment. The left panel shows the RMSE on retained positions; the right panel shows the corresponding non-predictable fraction. Pointwise curves are shown for four oracle thresholds. The black curve is the best block-wise result among the executed search runs. Faint traces show the unsmoothed lead-time summaries and solid curves show the corresponding smoothed trends.}
\label{fig:greedy-vs-daemon}
\end{figure*}

Figure~\ref{fig:greedy-vs-daemon} compares the pointwise search at $\varepsilon_{\mathrm{orc}}\in\{0.005,0.01,0.02,0.05\}$ with the block-wise search used in the existence experiment. Stricter pointwise thresholds retain small errors only while the non-predictable fraction rapidly approaches one. Looser thresholds delay this coverage collapse but admit larger errors. The block-wise search maintains a skipped fraction close to that imposed by the mask sampler while preserving a low conditional error over the full examined range. The comparison supports the use of block-wise mask search as the oracle construction in the main analysis.

\subsection{Weighted aggregation in the oracle ensemble}
\label{app:oracle-ensemble-aggregation}

Fix a starting position $t$ and the sampled active sets and candidate multisets from Sec.~\ref{subsec:ensemble-partial-trajectories}. At lead time $h$, let
\begin{equation}
\mathcal K_{t,h} = \left\{ k\in\{1,\ldots,\Nens\}: t+h\in A_t^{(k)} \right\}
\end{equation}
be the indices of active partial trajectories, and write $S_{t,h}^{(k)}=\widehat S_{t+h,A_t^{(k)}}$. Each $S_{t,h}^{(k)}$ is a multiset, so its cardinality and all sums retain pattern-level multiplicities.

The total candidate mass available at the current lead time is
\begin{equation}
Z_{t,h}^{\mathrm{orc}} = \sum_{k\in\mathcal K_{t,h}} w_{t,k}^{\mathrm{orc}}|S_{t,h}^{(k)}|.
\end{equation}
For $Z_{t,h}^{\mathrm{orc}}>0$, define
\begin{equation}
c_{t,h}^{\mathrm{orc}} = \frac{ \displaystyle \sum_{k\in\mathcal K_{t,h}} w_{t,k}^{\mathrm{orc}} \sum_{v\in S_{t,h}^{(k)}}v }
{ Z_{t,h}^{\mathrm{orc}} }
\end{equation}
and
\begin{equation}
d_{t,h}^{\mathrm{orc}} = \frac{ \displaystyle \sum_{k\in\mathcal K_{t,h}} w_{t,k}^{\mathrm{orc}} \sum_{v\in S_{t,h}^{(k)}} |v-c_{t,h}^{\mathrm{orc}}| }
{ Z_{t,h}^{\mathrm{orc}} }.
\end{equation}
The predictability decision is
\begin{equation}
\zeta_{\mathrm{orc}}(t,h;\teps)=1 \quad\Longleftrightarrow\quad Z_{t,h}^{\mathrm{orc}}>0 \ \text{and}\ d_{t,h}^{\mathrm{orc}}<\teps.
\end{equation}
Otherwise $\zeta_{\mathrm{orc}}=0$. At a predictable point,
\begin{equation}
\widehat y_{t+h}^{\mathrm{orc}}=c_{t,h}^{\mathrm{orc}}.
\end{equation}
Division by $Z_{t,h}^{\mathrm{orc}}$ renormalizes the trajectory weights over the candidate occurrences available at the current lead time. An inactive trajectory contributes no mass. All candidates from an active trajectory inherit the same trajectory-rank weight.

For $\gorc=0$, the rank weights are uniform and the weighted center and dispersion reduce to the ensemble definitions $c_{t,h}$ and $d_{t,h}$. The main comparison instead uses $\gorc=0.04$ and applies the same threshold selected for the ensemble without re-estimation.

\section{Local-predictor selection}
\label{app:predictor-selection}

For a pattern $\alpha$, the closest-motifs operator uses
\begin{equation}
\mathcal J_{\varepsilon}^{(\alpha)}(x) =
\left\{ m:\|x-x_m^{(\alpha)}\|_2\le\varepsilon \right\}
\end{equation}
and returns the target multiset
\begin{equation}
\mathcal M_\alpha(x) = \left\{ u_m^{(\alpha)}: m\in\mathcal J_{\varepsilon}^{(\alpha)}(x) \right\}.
\end{equation}
When the neighbourhood is empty, the operator uses the target of the nearest training context. We select the radius on the validation segment separately for each pair $(L,K_{\max})$.

The decision-tree predictor is an unrestricted regression tree with minimum leaf size one. The random forest is ten unrestricted trees with the same minimum leaf size. Both are deterministic at inference time and contribute singleton candidate multisets for each pattern.

We evaluate the complete grid on the Lorenz validation segment. Tables~\ref{tab:rmse_tree_grid}, \ref{tab:rmse_forest_grid}, and~\ref{tab:rmse_motifs_grid} report recursive-baseline RMSE at discrete leads $h\in\{1,10,30,50\}$, corresponding to $\Lambda_h\approx\{0.1,0.9,2.7,4.6\}$ Lyapunov times. Table~\ref{tab:best_eps_motifs} lists the validation-selected motif radii.

\begin{figure}[t]
    \centering
    \includegraphics[width=\columnwidth]
    {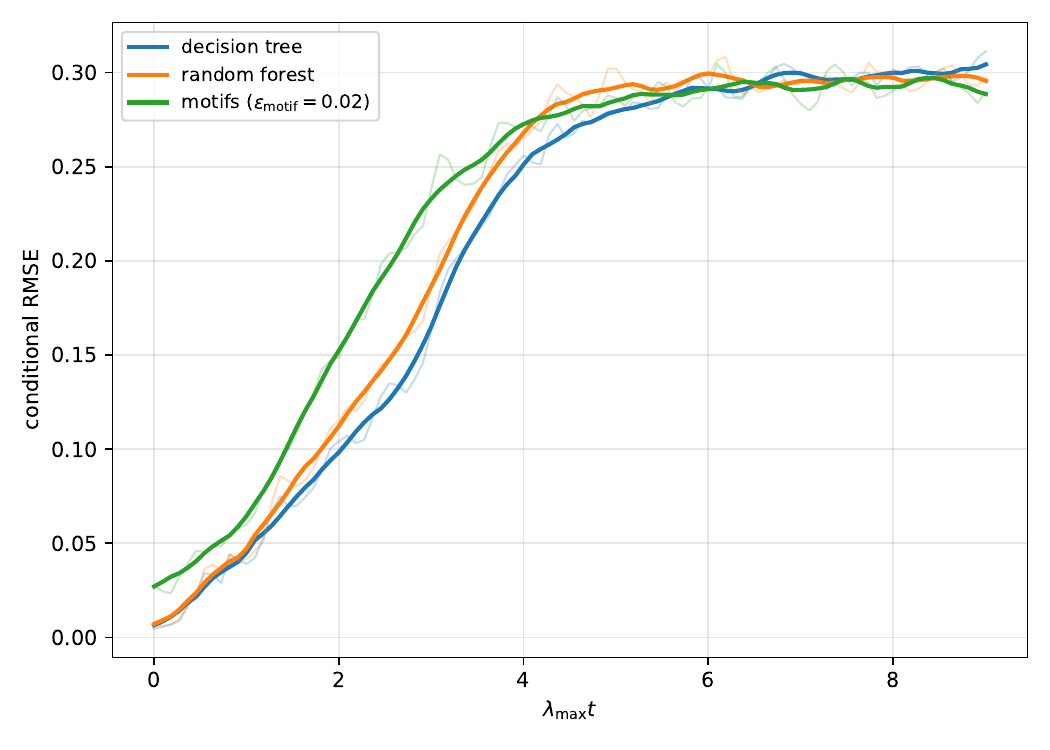}
    \caption{Recursive-baseline RMSE on the Lorenz validation segment for the decision-tree, random-forest, and closest-motifs predictors at $L=5$ and $K_{\max}=5$. The closest-motifs curve uses $\varepsilon_{\mathrm{motif}}=0.02$. Thin lines show the raw lead-time RMSE and thick lines its smoothed trend; the horizontal axis gives lead time in Lyapunov units.}
    \label{fig:local-predictor-comparison}
\end{figure}

\begin{table}[t]
\centering
\caption{Recursive-baseline RMSE of the decision-tree predictor for the tested
values of $L$ and $K_{\max}$. Columns report the discrete lead $h$ and, on the
second line, the corresponding Lorenz lead time $\Lambda_h$. Boldface marks
the lowest RMSE within each lead-time column.}
\label{tab:rmse_tree_grid}
\begingroup
\footnotesize
\setlength{\tabcolsep}{2.5pt}
\renewcommand{\arraystretch}{0.98}
\begin{tabular}{@{}ccrrrr@{}}
\toprule
$L$ & $K_{\max}$
& \multicolumn{1}{c}{$h=1$}
& \multicolumn{1}{c}{$h=10$}
& \multicolumn{1}{c}{$h=30$}
& \multicolumn{1}{c}{$h=50$} \\
&
& \multicolumn{1}{c}{$\Lambda_h\!\approx\!0.1$}
& \multicolumn{1}{c}{$\Lambda_h\!\approx\!0.9$}
& \multicolumn{1}{c}{$\Lambda_h\!\approx\!2.7$}
& \multicolumn{1}{c}{$\Lambda_h\!\approx\!4.6$} \\
\midrule
\multirow{9}{*}{4} & 3 & 0.0047 & 0.0661 & 0.2204 & 0.2879 \\
& 4 & 0.0052 & 0.0706 & 0.2320 & 0.2805 \\
& 5 & 0.0063 & 0.0577 & 0.2177 & 0.2789 \\
& 6 & 0.0090 & 0.0523 & 0.1968 & 0.2833 \\
& 7 & 0.0151 & 0.0498 & 0.1970 & 0.2655 \\
& 8 & 0.0227 & 0.0552 & 0.1928 & 0.2507 \\
& 9 & 0.0295 & 0.0570 & 0.1814 & 0.2393 \\
& 10 & 0.0350 & 0.0560 & 0.1771 & 0.2339 \\
& 11 & 0.0394 & 0.0591 & 0.1824 & \textbf{0.2280} \\
\addlinespace
\multirow{5}{*}{5} & 3 & 0.0035 & 0.0550 & 0.1573 & 0.2722 \\
& 4 & 0.0034 & 0.0493 & 0.1382 & 0.2757 \\
& 5 & 0.0046 & 0.0441 & 0.1339 & 0.2653 \\
& 6 & 0.0079 & 0.0381 & 0.1280 & 0.2662 \\
& 7 & 0.0146 & 0.0396 & 0.1415 & 0.2725 \\
\addlinespace
\multirow{4}{*}{6} & 3 & \textbf{0.0033} & 0.0439 & 0.1269 & 0.2694 \\
& 4 & 0.0033 & 0.0354 & 0.1166 & 0.2585 \\
& 5 & 0.0047 & \textbf{0.0350} & \textbf{0.1143} & 0.2607 \\
& 6 & 0.0082 & 0.0364 & 0.1172 & 0.2619 \\
\bottomrule
\end{tabular}
\endgroup
\end{table}

\begin{table}[t]
\centering
\caption{Recursive-baseline RMSE of the random-forest predictor for the tested
values of $L$ and $K_{\max}$. Columns report the discrete lead $h$ and, on the
second line, the corresponding Lorenz lead time $\Lambda_h$. Boldface marks
the lowest RMSE within each lead-time column.}
\label{tab:rmse_forest_grid}
\begingroup
\footnotesize
\setlength{\tabcolsep}{2.5pt}
\renewcommand{\arraystretch}{0.98}
\begin{tabular}{@{}ccrrrr@{}}
\toprule
$L$ & $K_{\max}$
& \multicolumn{1}{c}{$h=1$}
& \multicolumn{1}{c}{$h=10$}
& \multicolumn{1}{c}{$h=30$}
& \multicolumn{1}{c}{$h=50$} \\
&
& \multicolumn{1}{c}{$\Lambda_h\!\approx\!0.1$}
& \multicolumn{1}{c}{$\Lambda_h\!\approx\!0.9$}
& \multicolumn{1}{c}{$\Lambda_h\!\approx\!2.7$}
& \multicolumn{1}{c}{$\Lambda_h\!\approx\!4.6$} \\
\midrule
\multirow{9}{*}{4} & 3 & 0.0048 & 0.0701 & 0.2231 & 0.2912 \\
& 4 & 0.0054 & 0.0763 & 0.2335 & 0.2842 \\
& 5 & 0.0070 & 0.0652 & 0.2319 & 0.2914 \\
& 6 & 0.0109 & 0.0593 & 0.2160 & 0.2839 \\
& 7 & 0.0178 & 0.0578 & 0.2146 & 0.2688 \\
& 8 & 0.0256 & 0.0619 & 0.2136 & 0.2458 \\
& 9 & 0.0327 & 0.0636 & 0.1991 & 0.2366 \\
& 10 & 0.0382 & 0.0632 & 0.1940 & 0.2273 \\
& 11 & 0.0429 & 0.0672 & 0.1964 & \textbf{0.2229} \\
\addlinespace
\multirow{5}{*}{5} & 3 & 0.0039 & 0.0566 & 0.1725 & 0.2881 \\
& 4 & 0.0039 & 0.0479 & 0.1565 & 0.2895 \\
& 5 & 0.0058 & 0.0444 & 0.1557 & 0.2893 \\
& 6 & 0.0100 & 0.0415 & 0.1520 & 0.2819 \\
& 7 & 0.0172 & 0.0425 & 0.1586 & 0.2735 \\
\addlinespace
\multirow{3}{*}{6} & 3 & \textbf{0.0037} & 0.0390 & 0.1335 & 0.2759 \\
& 4 & 0.0039 & \textbf{0.0305} & \textbf{0.1323} & 0.2747 \\
& 5 & 0.0060 & 0.0369 & 0.1413 & 0.2769 \\
\bottomrule
\end{tabular}
\endgroup
\end{table}

\begin{table}[t]
\centering
\caption{Recursive-baseline RMSE of the closest-motifs predictor for the tested
values of $L$ and $K_{\max}$, using the validation-selected
$\varepsilon_{\mathrm{motif}}$ for each pair. Columns report the discrete lead
$h$ and, on the second line, the corresponding Lorenz lead time $\Lambda_h$.
Boldface marks the lowest RMSE within each lead-time column.}
\label{tab:rmse_motifs_grid}
\begingroup
\footnotesize
\setlength{\tabcolsep}{2.5pt}
\renewcommand{\arraystretch}{0.98}
\begin{tabular}{@{}ccrrrr@{}}
\toprule
$L$ & $K_{\max}$
& \multicolumn{1}{c}{$h=1$}
& \multicolumn{1}{c}{$h=10$}
& \multicolumn{1}{c}{$h=30$}
& \multicolumn{1}{c}{$h=50$} \\
&
& \multicolumn{1}{c}{$\Lambda_h\!\approx\!0.1$}
& \multicolumn{1}{c}{$\Lambda_h\!\approx\!0.9$}
& \multicolumn{1}{c}{$\Lambda_h\!\approx\!2.7$}
& \multicolumn{1}{c}{$\Lambda_h\!\approx\!4.6$} \\
\midrule
\multirow{9}{*}{4} & 3 & 0.0077 & 0.1026 & 0.2655 & 0.2933 \\
& 4 & 0.0127 & 0.0994 & 0.2805 & 0.2880 \\
& 5 & 0.0238 & 0.1057 & 0.2742 & 0.2814 \\
& 6 & 0.0362 & 0.0856 & 0.2645 & 0.2729 \\
& 7 & 0.0483 & 0.0736 & 0.2614 & 0.2602 \\
& 8 & 0.0495 & 0.0619 & 0.2353 & 0.2622 \\
& 9 & 0.0608 & 0.0756 & 0.2414 & 0.2516 \\
& 10 & 0.0639 & 0.0777 & 0.2397 & 0.2523 \\
& 11 & 0.0678 & 0.0826 & 0.2357 & \textbf{0.2453} \\
\addlinespace
\multirow{5}{*}{5} & 3 & \textbf{0.0063} & 0.0632 & 0.2040 & 0.2786 \\
& 4 & 0.0102 & 0.0664 & 0.2060 & 0.2838 \\
& 5 & 0.0276 & 0.0542 & 0.2047 & 0.2827 \\
& 6 & 0.0404 & \textbf{0.0481} & 0.1875 & 0.2753 \\
& 7 & 0.0503 & 0.0494 & 0.1843 & 0.2764 \\
\addlinespace
\multirow{4}{*}{6} & 3 & 0.0080 & 0.0560 & 0.1972 & 0.2948 \\
& 4 & 0.0170 & 0.0487 & 0.1888 & 0.2787 \\
& 5 & 0.0362 & 0.0520 & 0.1801 & 0.2729 \\
& 6 & 0.0461 & 0.0501 & \textbf{0.1776} & 0.2705 \\
\bottomrule
\end{tabular}
\endgroup
\end{table}

\begin{table}[t]
\centering
\caption{Validation-selected closest-motifs radius
$\varepsilon_{\mathrm{motif}}$ for each tested pair $(L,K_{\max})$.}
\label{tab:best_eps_motifs}
\begingroup
\footnotesize
\setlength{\tabcolsep}{5pt}
\renewcommand{\arraystretch}{0.98}
\begin{tabular}{@{}ccc@{}}
\toprule
$L$ & $K_{\max}$ & $\varepsilon_{\mathrm{motif}}$ \\
\midrule
\multirow{9}{*}{4} & 3 & 0.0100 \\
& 4 & 0.0150 \\
& 5 & 0.0200 \\
& 6 & 0.0200 \\
& 7 & 0.0200 \\
& 8 & 0.0150 \\
& 9 & 0.0200 \\
& 10 & 0.0200 \\
& 11 & 0.0200 \\
\addlinespace
\multirow{5}{*}{5} & 3 & 0.0100 \\
& 4 & 0.0150 \\
& 5 & 0.0200 \\
& 6 & 0.0200 \\
& 7 & 0.0200 \\
\addlinespace
\multirow{4}{*}{6} & 3 & 0.0200 \\
& 4 & 0.0150 \\
& 5 & 0.0200 \\
& 6 & 0.0200 \\
\bottomrule
\end{tabular}
\endgroup
\end{table}

\FloatBarrier

No single configuration minimizes RMSE at every reported lead time. Larger values of $L$ and $K_{\max}$ improve some intermediate-horizon entries but also increase the number of generalized patterns and the fitting cost. The selected configuration $L=5$, $K_{\max}=5$ retains short- and intermediate-horizon accuracy close to the best grid values while keeping the pattern family smaller than the largest examined families. Under this common configuration, the decision tree gives the smallest one-step RMSE among the tested predictor classes, so we retain it for all subsequent experiments.

\section{Predictor-class robustness}
\label{app:predictor-robustness}

The structural experiments use an unrestricted regression tree as the local candidate operator. This appendix repeats the partial-ensemble construction on the Lorenz validation segment with two other local predictors: a local linear predictor and the closest-motifs operator.

The local linear predictor fits, for each pattern $\alpha$, a linear regression on the pattern contexts of the training segment, without regularization. Like the tree, it contributes a singleton candidate multiset per pattern. The closest-motifs operator uses $\varepsilon_{\mathrm{motif}}=0.02$, the radius selected in Appendix~\ref{app:predictor-selection}. Both experiments use the pattern and mask configuration of Sec.~\ref{subsec:shared-configuration}, $\Nens=100$ partial trajectories per starting position, and 1\,000 starting positions, evaluated up to 13.7 Lyapunov times. The threshold sweep and Pareto extraction follow Sec.~\ref{subsec:evaluation-metrics}.

\begin{figure*}[tbp]
  \centering
  \includegraphics[width=0.95\textwidth]
  {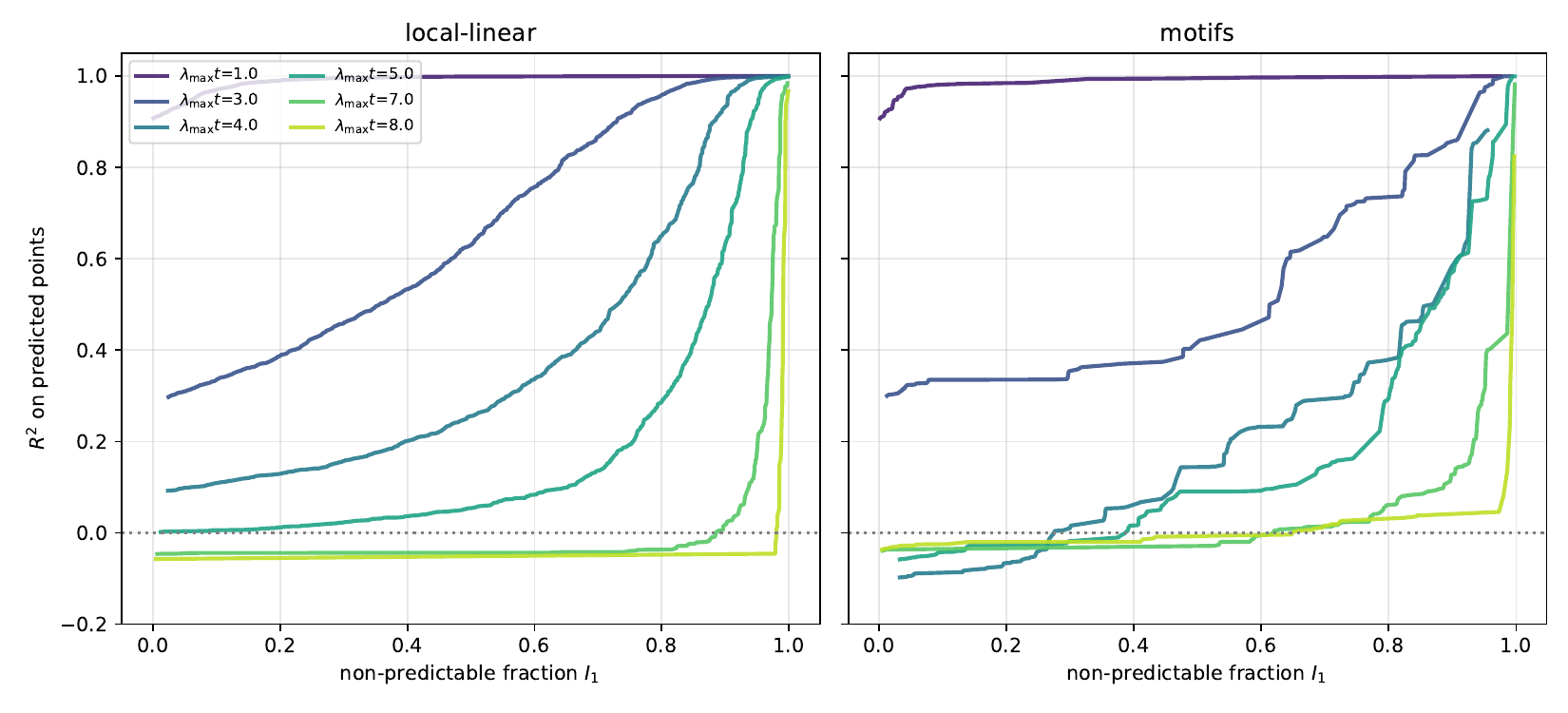}
  \caption{$R^2$--non-predictability trade-offs of the partial trajectory ensemble with the local linear (left) and closest-motifs ($\varepsilon_{\mathrm{motif}}=0.02$, right) predictors on the Lorenz validation segment, at 1\,000 starting positions with $\Nens=100$. Each curve contains nondominated points from the threshold sweep; colours identify lead times in Lyapunov units.}
  \label{fig:predictor-robustness-pareto}
\end{figure*}

Figure~\ref{fig:predictor-robustness-pareto} shows the resulting $R^2$--non-predictability fronts. Both predictors reproduce the structure of the tree-based fronts: short-lead fronts remain predominantly concave, long-lead fronts become predominantly convex, and only strongly selected subsets reach positive conditional $R^2$ at long lead times. The transition between the two regimes occurs earlier for these weaker local predictors than for the tree, near 3--4 Lyapunov times instead of between 4 and 5, consistent with their larger baseline errors. The location $\Lambda^\ast$ therefore depends on the local predictor, while the transition itself does not.

\begin{figure}[tbp]
  \centering
  \includegraphics[width=0.85\columnwidth]
  {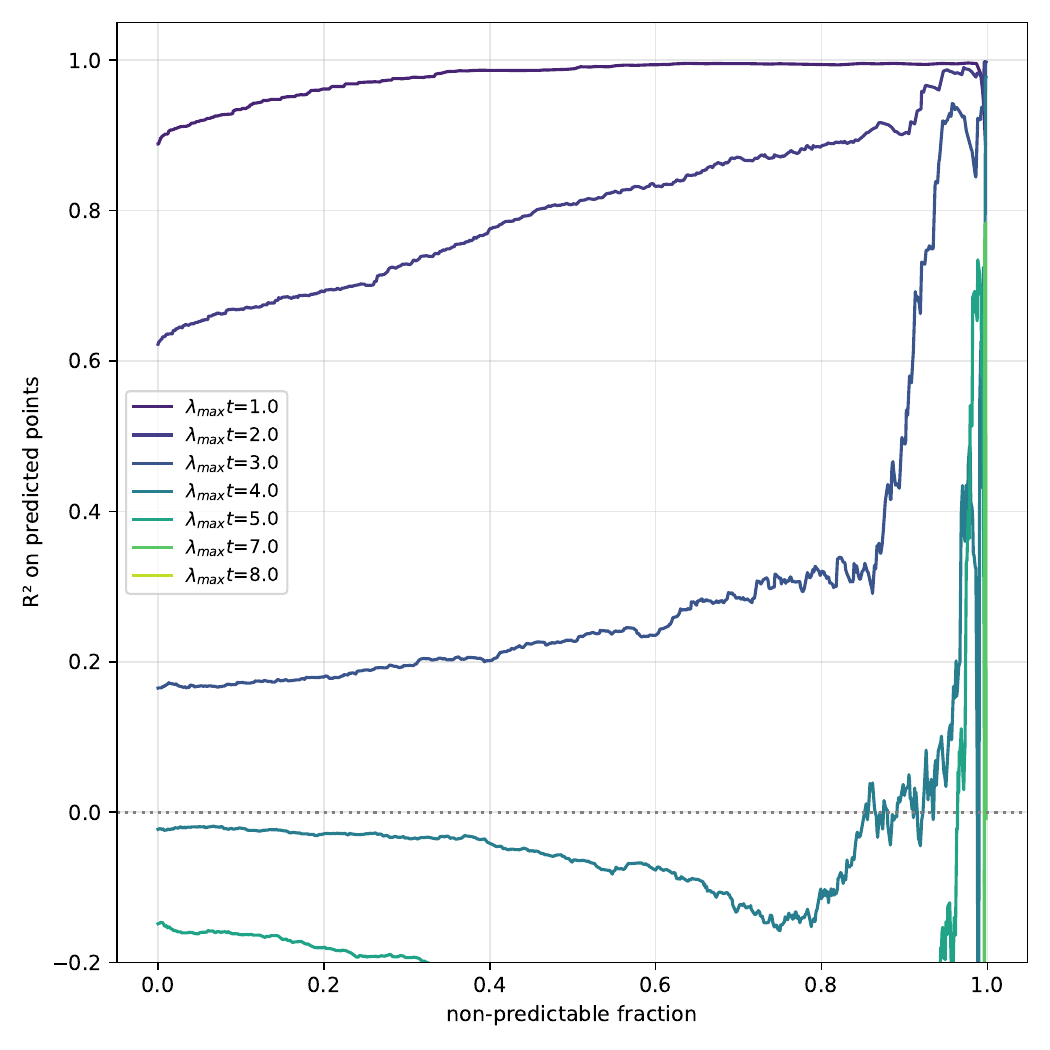}
  \caption{$R^2$--non-predictability trade-off of the perturbed generator with the local linear predictor at the tuned perturbation scale $\drel=0.010$ on the Lorenz validation segment. Colours identify lead times in Lyapunov units. This panel is the perturbed counterpart to the local-linear panel of Fig.~\ref{fig:predictor-robustness-pareto}.}
  \label{fig:predictor-robustness-perturbed-loclin}
\end{figure}

The perturbed generator reproduces the same qualitative $R^2$--non-predictability structure at the tuned scale $\drel=0.010$ when the local predictor is the local linear one (Fig.~\ref{fig:predictor-robustness-perturbed-loclin}), matching the partial-ensemble front of the corresponding panel of Fig.~\ref{fig:predictor-robustness-pareto}.

Front levels are not compared across predictors at large $I_1$. At extreme selectivity the accepted subsets contain few starting positions, and the conditional metric becomes unstable.

\section{Supplementary ensemble results}
\label{app:ensemble-results}

\subsection{RMSE-based Pareto fronts}
\label{app:rmse-pareto}

\begin{figure*}[tbp]
\centering
\begin{minipage}[t]{0.46\textwidth}
\centering
\includegraphics[width=\linewidth]
{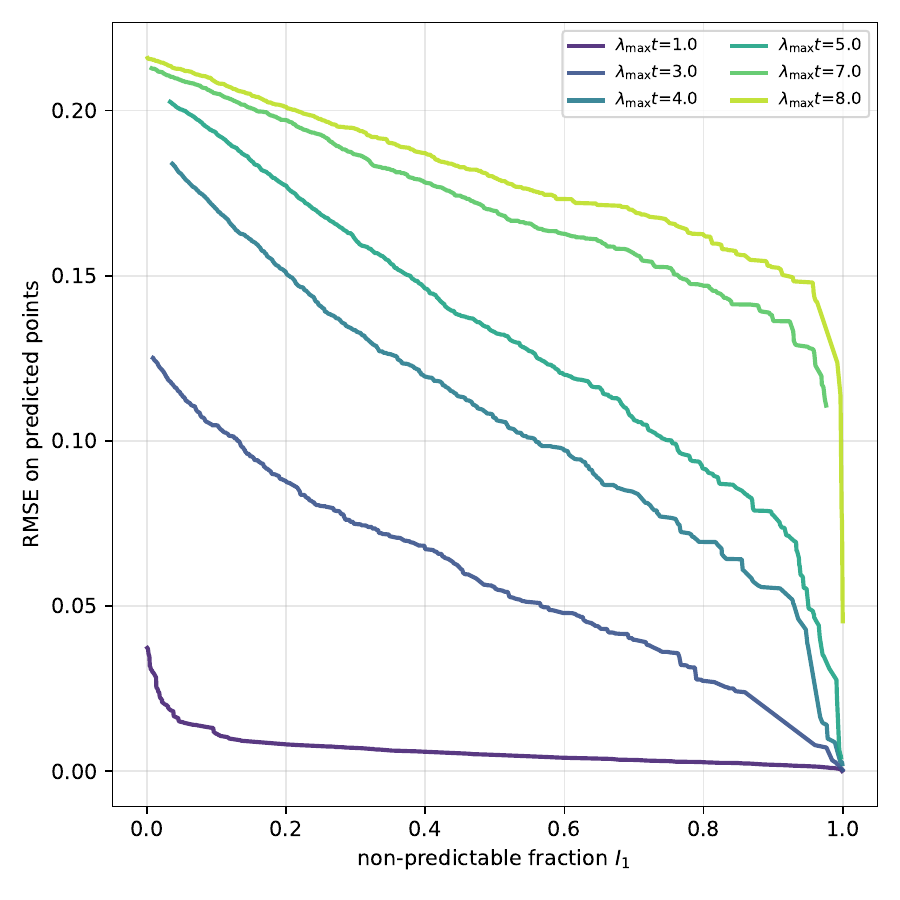}
\caption*{(a) Ensemble.}
\end{minipage}
\hfill
\begin{minipage}[t]{0.46\textwidth}
\centering
\includegraphics[width=\linewidth]
{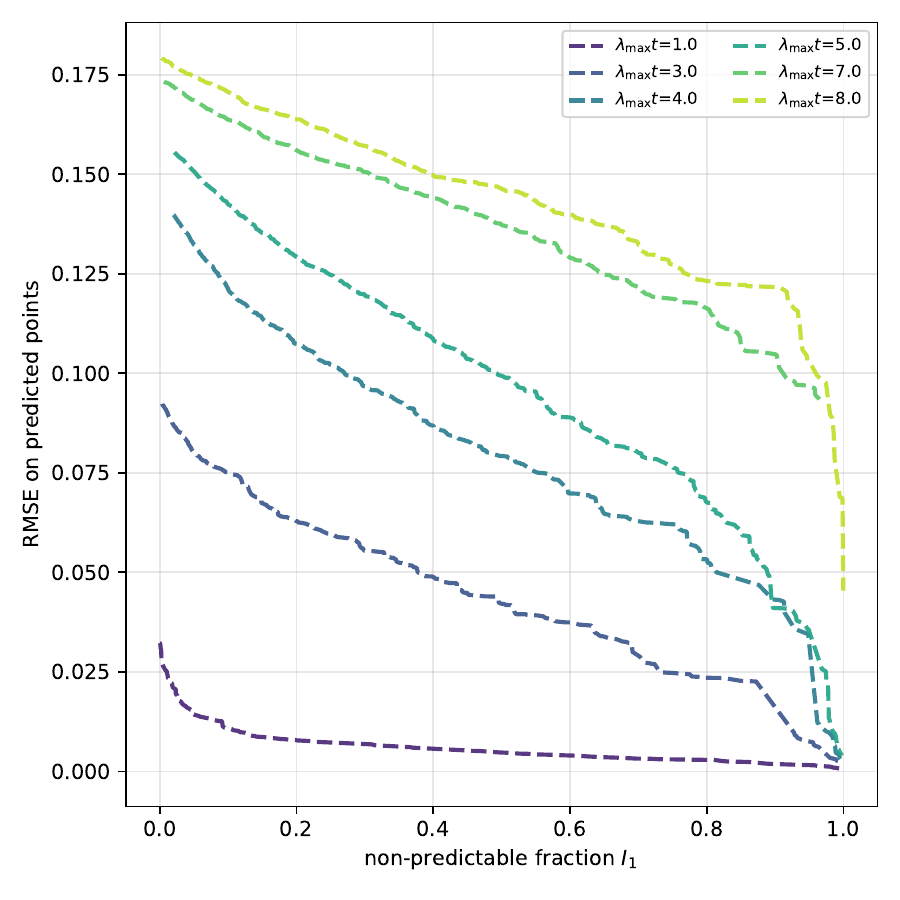}
\caption*{(b) Oracle ensemble.}
\end{minipage}
\caption{RMSE-based Pareto fronts for the ensemble and oracle ensemble on the Lorenz validation segment. The Pareto construction minimizes RMSE and the non-predictable fraction and removes dominated operating points separately for each lead time and reduction rule. The displayed lead times match the representative $R^2$ fronts in the main text, with units of Lyapunov time.}
\label{fig:rmse-pareto}
\end{figure*}

The main analysis uses conditional $R^2$ to evaluate performance relative to the variance of the retained targets. The complementary error-scale representation appears in Fig.~\ref{fig:rmse-pareto}. Each curve contains all nondominated operating points generated by the threshold sweep. We apply no additional coverage prefilter.

At short lead times, the ensemble fronts reach low RMSE while retaining a large fraction of starting positions. At later lead times, comparable errors require more selective operating points. The oracle ensemble shifts the fronts toward lower RMSE and lower non-predictability, showing how trajectory ranking changes the same sampled candidate pool.

\subsection{Partial versus perturbed Pareto fronts}
\label{app:partial-perturbed-fronts}

\begin{figure*}[tbp]
  \centering
  \includegraphics[width=0.92\textwidth]
  {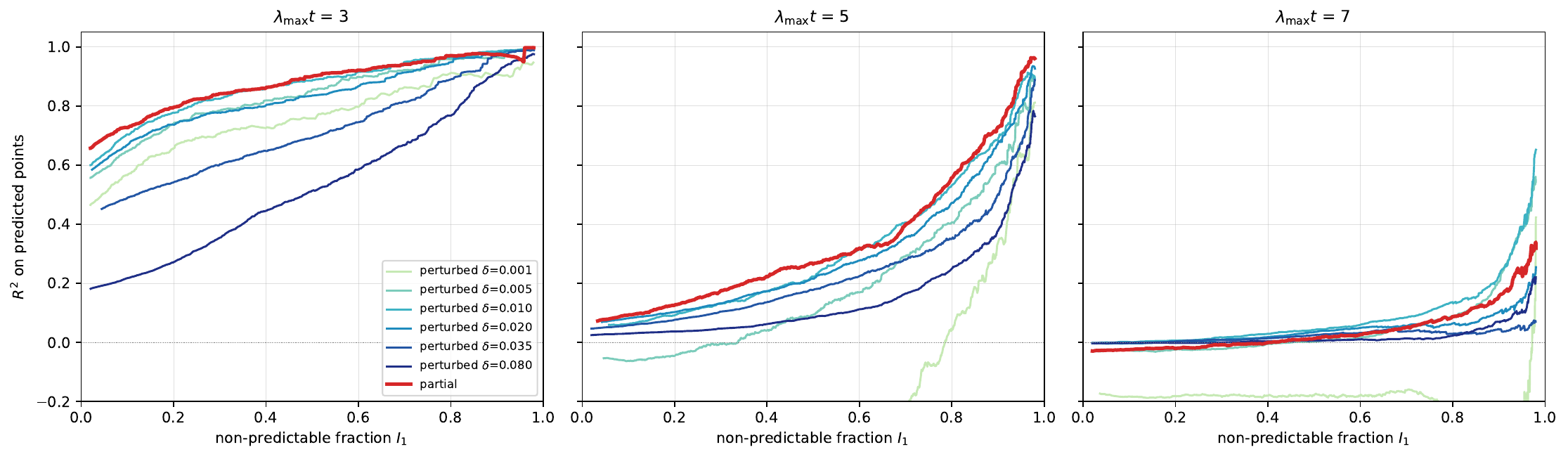}
  \caption{Pareto comparison between the partial trajectory ensemble and perturbed generators on the Lorenz series. Each point corresponds to a dispersion threshold and reports conditional $R^2$ against the non-predictable fraction. Perturbed curves are shown for $\drel\in\{0.001,0.005,0.010,0.020,0.035,0.080\}$. The partial trajectory ensemble is shown as one curve because it has no noise scale.}
  \label{fig:partial-perturbed-pareto}
\end{figure*}

Figure~\ref{fig:partial-perturbed-pareto} reports the complete $R^2$--non-predictability fronts behind the fixed-coverage comparison of Sec.~\ref{subsec:results-partial-perturbed}, with one curve per perturbation scale. At $\Lambda_h\approx3.0$, the best perturbed curves and the partial front are nearly tied over most of the coverage range. At $\Lambda_h\approx5.0$, the partial front lies above every perturbed curve over the displayed range, not only at $I_1=0.5$. At $\Lambda_h\approx7.0$, the best perturbed curves lie slightly above the partial front near the reported operating point.

\subsection{Sensitivity to oracle trajectory weights}
\label{app:oracle-weight-sensitivity}

The principal oracle ensemble uses $\gorc=0.04$. Increasing $\gorc$ concentrates more mass on the highest-ranked partial trajectories while leaving the starting positions, masks, candidate multisets, and predictability threshold unchanged.

\begin{figure*}[tbp]
\centering
\begin{minipage}[t]{0.46\textwidth}
\centering
\includegraphics[width=\linewidth]
{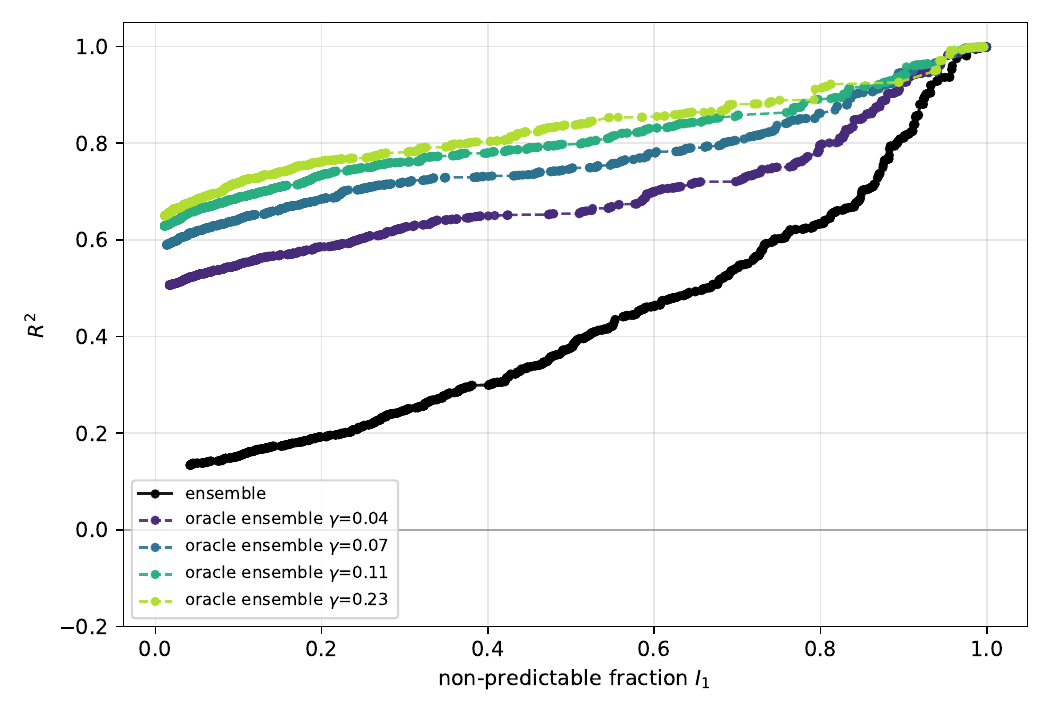}
\caption*{(a) 4.6 Lyapunov times.}
\end{minipage}
\hfill
\begin{minipage}[t]{0.46\textwidth}
\centering
\includegraphics[width=\linewidth]
{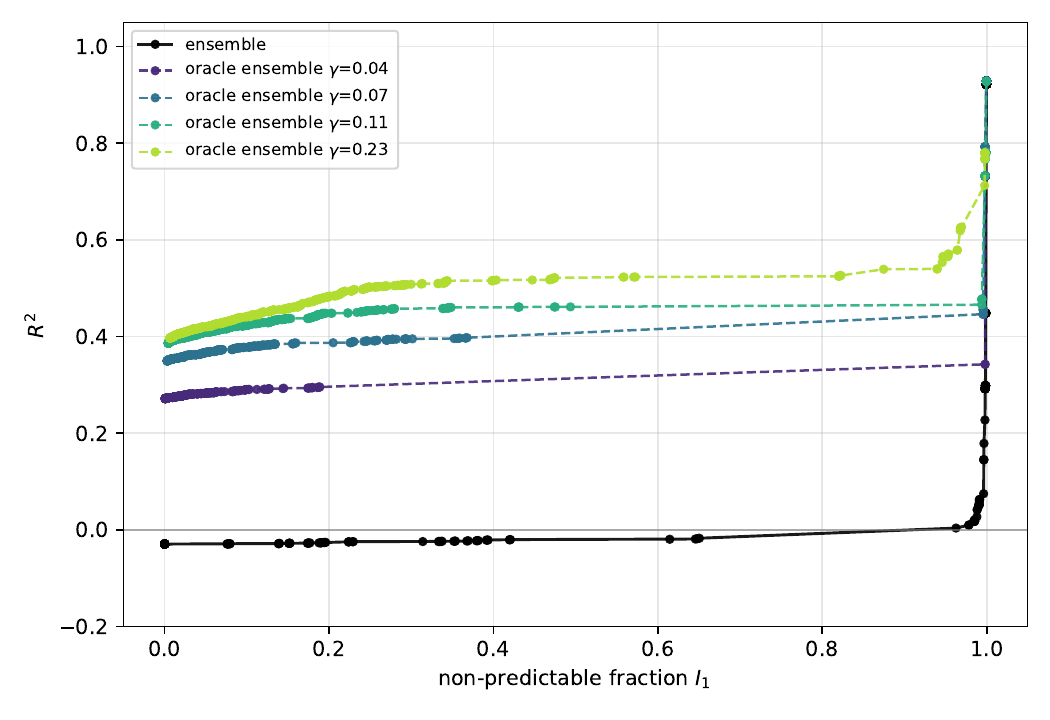}
\caption*{(b) 9.1 Lyapunov times.}
\end{minipage}
\caption{Sensitivity of the oracle-ensemble trade-off to the trajectory-rank concentration parameter. The ensemble is the unweighted reference; the remaining curves apply alternative values of $\gorc$ to the same sampled trajectory pools.}
\label{fig:oracle-weight-sensitivity}
\end{figure*}

Figure~\ref{fig:oracle-weight-sensitivity} shows the resulting $R^2$--non-predictability trade-offs at 4.6 and 9.1 Lyapunov times. Stronger concentration generally increases conditional $R^2$, although the gain is not uniform across the full coverage range. We use the principal value as a fixed moderate concentration rather than selecting it separately on the test outcomes.

\subsection{Oracle-ensemble scroll diagnostics}
\label{app:oracle-scroll}

The scroll analysis of Sec.~\ref{subsec:results-topological-transition} compares the two generators. The oracle-ensemble reduction of the partial mask pools provides a complementary reference at the same two operating points. Both reductions use identical starting positions, mask pools, and thresholds, and differ only in the trajectory-rank weights.

\begin{figure*}[tbp]
    \centering
    \includegraphics[width=0.8\textwidth]
    {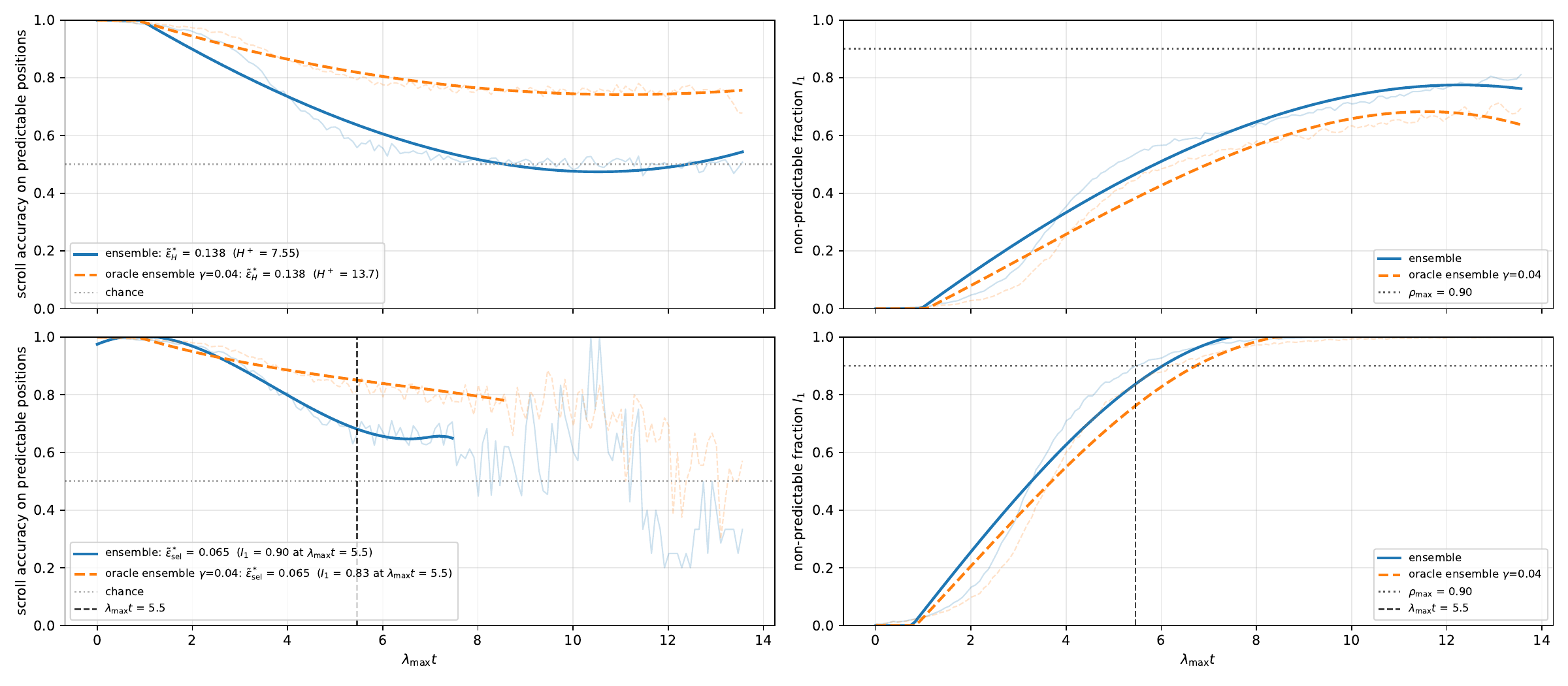}
    \captionsetup{width=0.96\textwidth}
    \caption{Scroll classification at two fixed operating points on 5\,000 Lorenz validation starting positions. Solid curves represent the ensemble; dashed curves represent the oracle ensemble with $\gorc=0.04$. Both reductions use identical starting positions, mask pools, and thresholds. The upper row uses $\teps_H^\ast=0.138$, selected by maximizing the positive-$R^2$ horizon under $\rho_{\max}=0.90$; the corresponding horizons are 7.55 Lyapunov times for the ensemble and 13.7 for the oracle ensemble. The lower row uses $\teps_{\mathrm{sel}}^\ast=0.065$, selected by maximizing conditional $R^2$ at $h_{\mathrm{sel}}=60$ ($\Lambda_{h_{\mathrm{sel}}}\approx5.5$ Lyapunov times). Left panels show scroll accuracy on predictable starting positions; right panels show the ensemble-level non-predictable fraction. Dotted lines mark chance accuracy and $\rho_{\max}$.}
    \label{fig:scroll-operating-points}
\end{figure*}

At the principal threshold $\teps_H^\ast=0.138$, ensemble scroll accuracy approaches chance after approximately 7--8 Lyapunov times, whereas the oracle ensemble remains above chance and retains a larger fraction of starting positions (Fig.~\ref{fig:scroll-operating-points}). The sampled pool contains long-range scroll information that the unweighted reduction does not fully recover. At the selective threshold $\teps_{\mathrm{sel}}^\ast=0.065$, chosen at 5.5 Lyapunov times, the retained subsets are more selective and have higher scroll accuracy. With the threshold fixed, scroll accuracy remains above chance at approximately 6.4--7.3 Lyapunov times, although the predictable fraction continues to decrease.

\section{Representative Lorenz predictions}
\label{app:lorenz-examples}

\begin{figure*}[tbp]
    \centering

    \begin{subfigure}{0.75\textwidth}
        \centering
        \includegraphics[width=\linewidth]
        {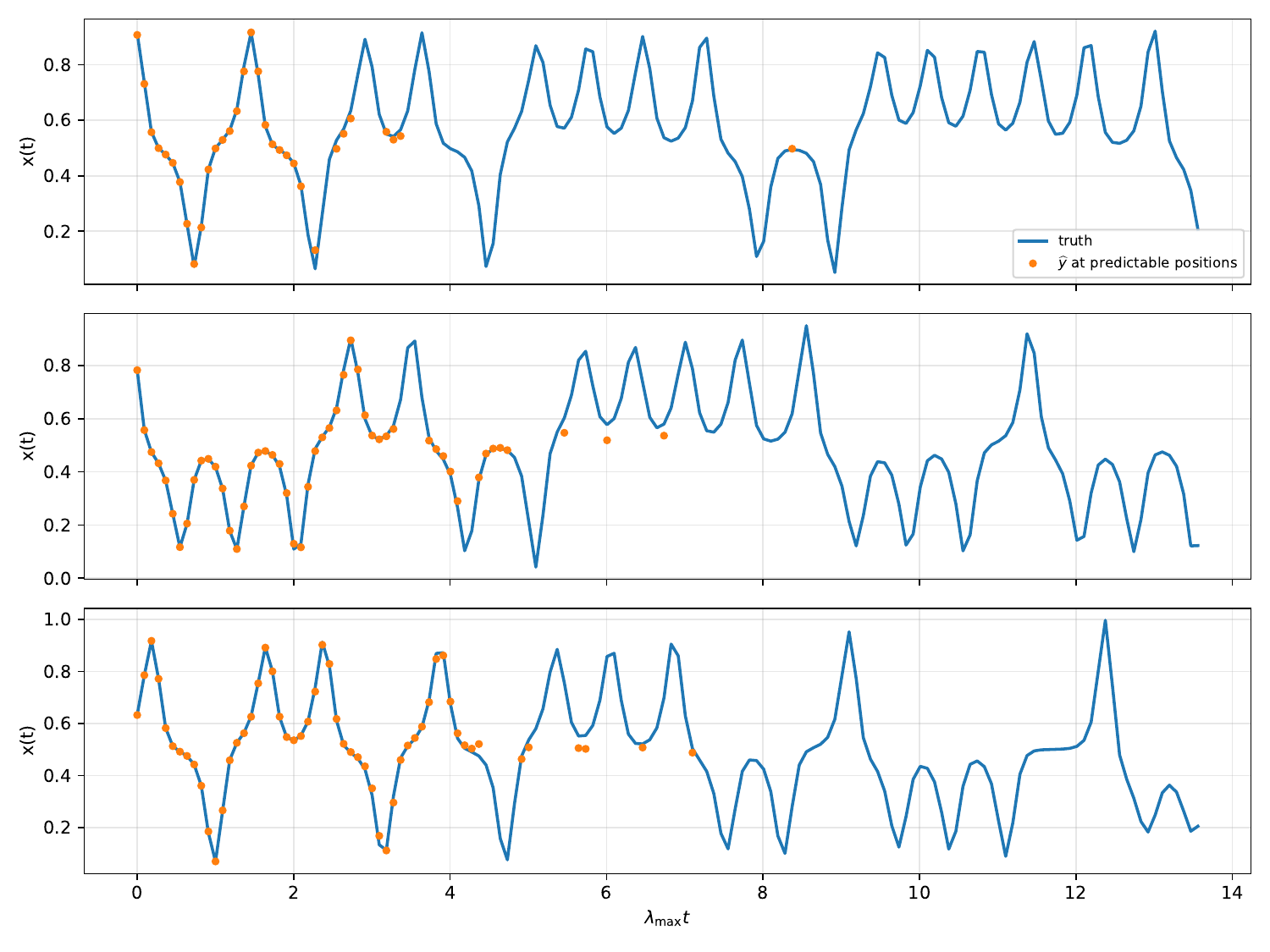}
        \caption{Ensemble.}
        \label{fig:representative-ensemble}
    \end{subfigure}

    \vspace{0.5em}

    \begin{subfigure}{0.75\textwidth}
        \centering
        \includegraphics[width=\linewidth]
        {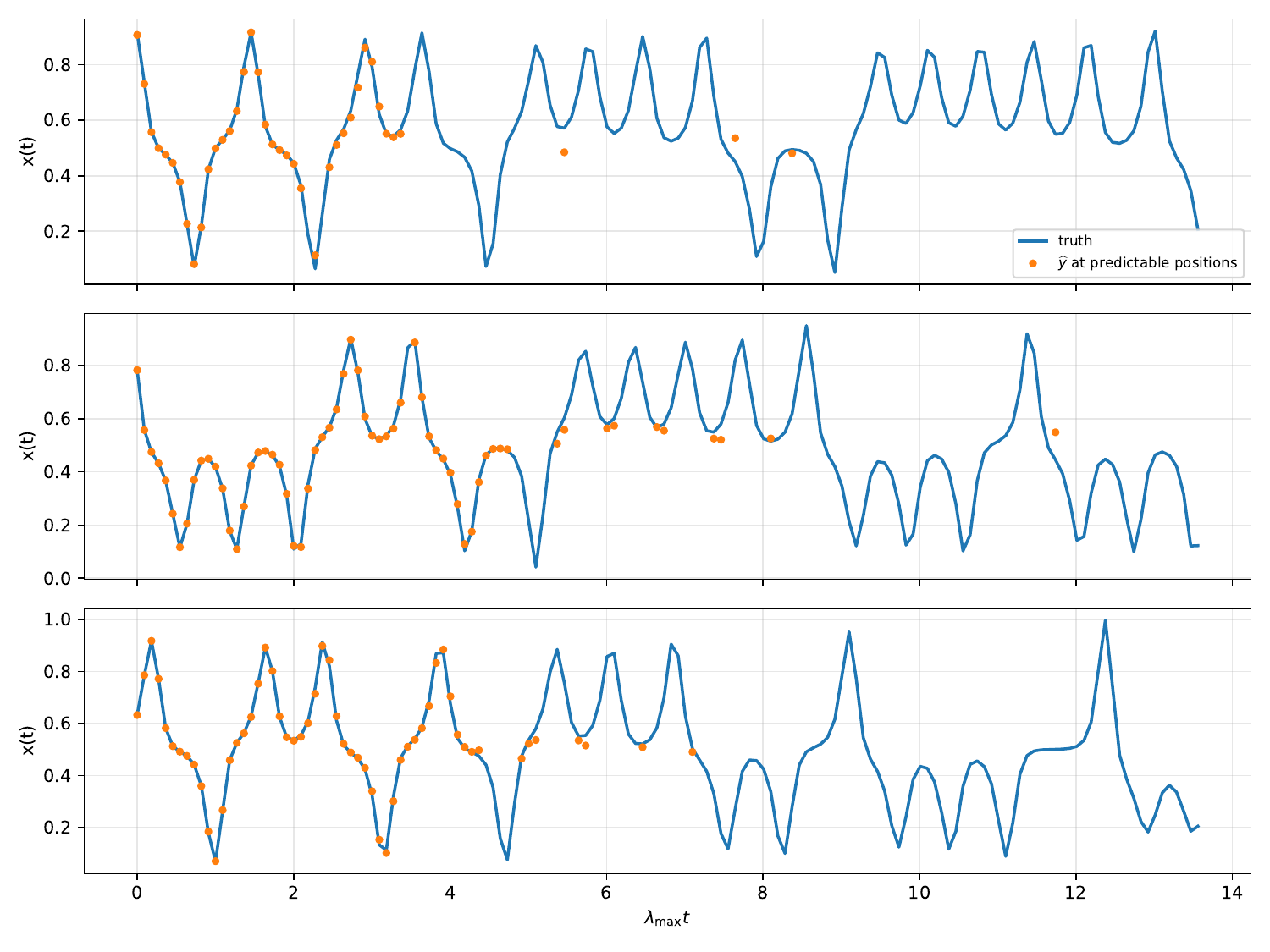}
        \caption{Oracle ensemble.}
        \label{fig:representative-oracle}
    \end{subfigure}

    \caption{Representative predictions at $\teps_{\mathrm{sel}}^{\ast}=0.065$, which the selection rule chooses at $h_{\mathrm{sel}}=60$ ($\Lambda_{h_{\mathrm{sel}}}\approx5.5$ Lyapunov times). The solid curves show the true globally normalized Lorenz series, and the markers show predictions at positions classified as predictable. Panels (a) and (b) use the same starting positions, sampled masks, and threshold. The oracle ensemble additionally applies trajectory-rank weighting with $\gorc=0.04$.}
    \label{fig:representative-forecasts}
\end{figure*}

Figure~\ref{fig:representative-forecasts} shows representative predictions at the selective operating point $\teps_{\mathrm{sel}}^{\ast}=0.065$, which the selection rule chooses at 5.5 Lyapunov times. The examples do not influence threshold selection. They illustrate how the aggregate metrics in Sec.~\ref{subsec:results-topological-transition} appear at individual starting positions.

In the ensemble examples (Fig.~\ref{fig:representative-ensemble}), early retained predictions follow the continuous trajectory closely. At later lead times, some retained values continue to identify the occupied scroll without reproducing its phase and amplitude accurately. This is the trajectory-level manifestation of the transition identified by the Pareto and correct-scroll diagnostics.

The oracle examples (Fig.~\ref{fig:representative-oracle}) use the same threshold and the principal rank concentration $\gorc=0.04$. Only the trajectory-rank weights differ from the ensemble reduction.

\section{Cross-system supplementary results}
\label{app:cross-system}

\begin{figure*}[t]
  \centering

  \begin{minipage}[t]{0.45\textwidth}
    \centering
    \includegraphics[width=\linewidth]
    {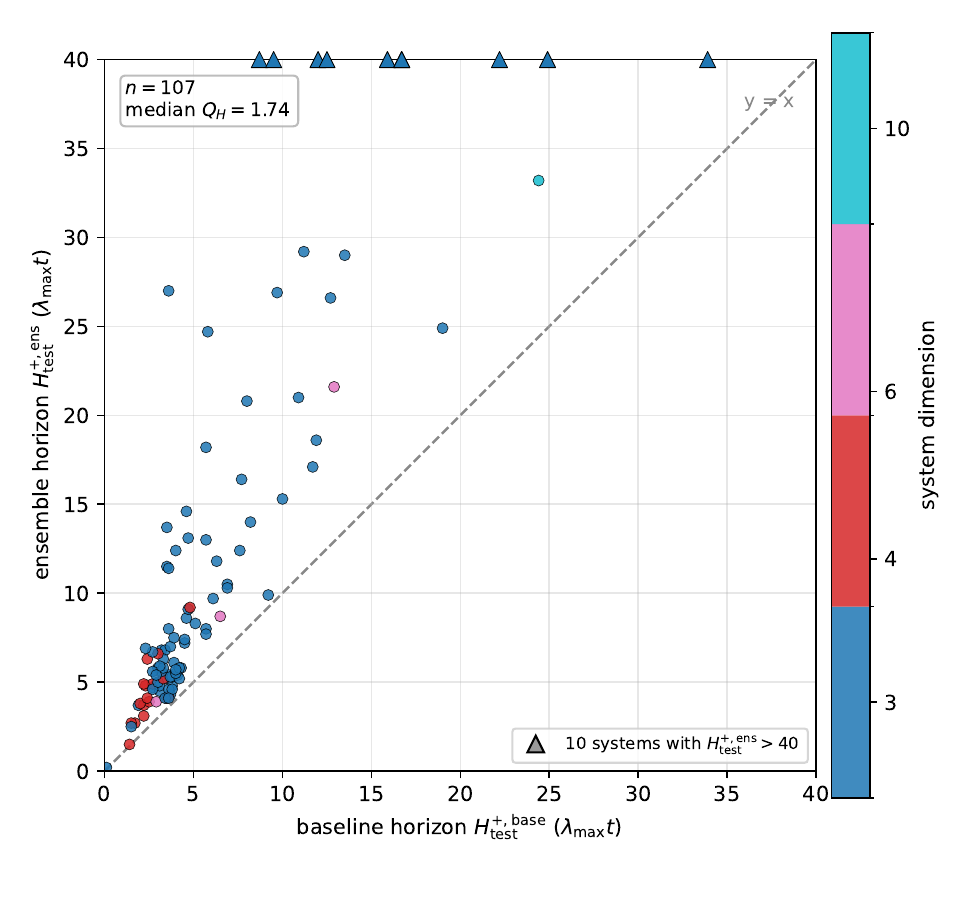}
    \caption*{(a) State dimension.}
  \end{minipage}
  \hfill
  \begin{minipage}[t]{0.45\textwidth}
    \centering
    \includegraphics[width=\linewidth]
    {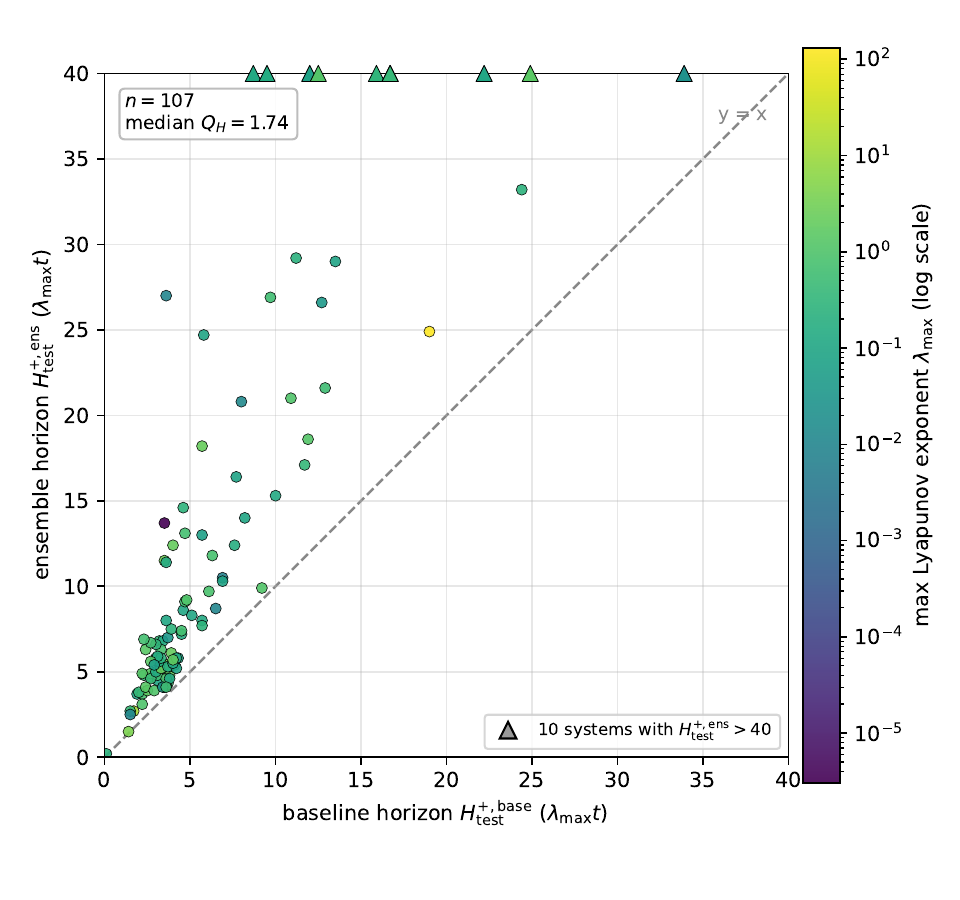}
    \caption*{(b) Largest Lyapunov exponent.}
  \end{minipage}

  \par\medskip

  \includegraphics[width=0.38\textwidth]
  {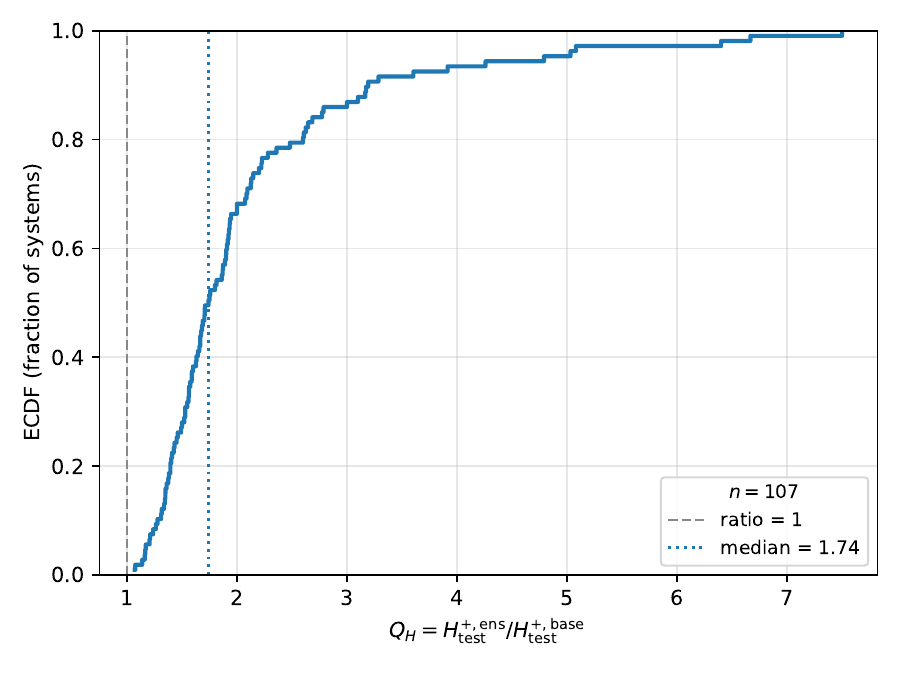}
  \caption*{(c) Empirical distribution of $Q_H$.}

  \caption{Cross-system positive-$R^2$ horizon diagnostics. Panels (a) and (b) compare the recursive-baseline and ensemble horizons, with colour indicating state dimension and the largest Lyapunov exponent, respectively. The dashed diagonal marks equal horizons, and triangles at the upper edge mark systems whose ensemble horizon exceeds 40 Lyapunov times. Panel (c) shows the empirical cumulative distribution of $Q_H=H_{\mathrm{test}}^{+,\mathrm{ens}}/ H_{\mathrm{test}}^{+,\mathrm{base}}$; the vertical reference marks $Q_H=1$. We compute horizons from the minimum componentwise $R^2$.}
  \label{fig:cross-system-diagnostics}
\end{figure*}

The scatter plot in Sec.~\ref{subsec:results-cross-system} compares absolute positive-$R^2$ horizons. Figure~\ref{fig:cross-system-diagnostics}(c) shows the empirical distribution of $Q_H$ [Eq.~\eqref{eq:qh}] across the retained systems.

Figure~\ref{fig:cross-system-diagnostics} colours the horizon comparison by state dimension and largest Lyapunov exponent. Systems of different dimensions appear both close to the diagonal and among the larger extensions. Extended horizons likewise occur across a broad range of largest Lyapunov exponents.

\subsection{Excluded systems}
\label{app:excluded-systems}

\begin{figure}[tbp]
  \centering
  \includegraphics[width=0.7\linewidth]
  {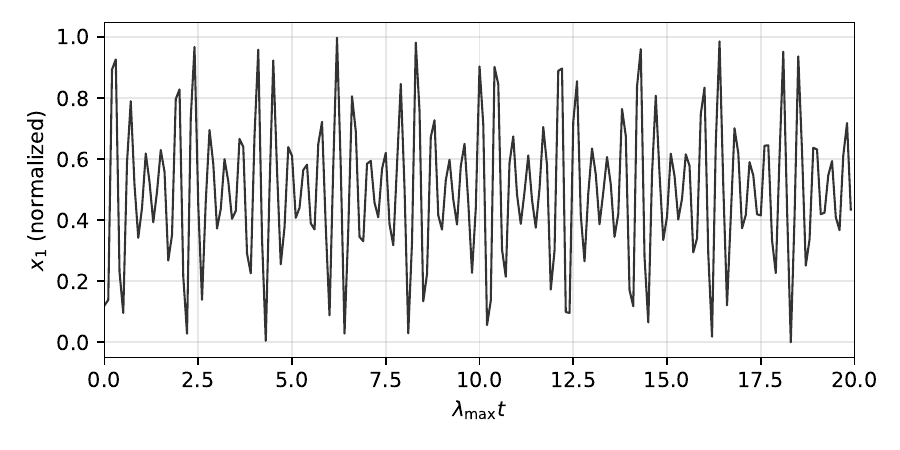}
  \caption{First state component of the quasi-periodic system \emph{SprottA} (Table~\ref{tab:excluded-systems}) over the first 20 Lyapunov times, showing the near-recurrent oscillation that keeps the recursive-baseline $R^2$ above zero beyond the 40-Lyapunov-time threshold.}
  \label{fig:excluded-example}
\end{figure}

Table~\ref{tab:excluded-systems} lists the systems excluded by the censoring criterion of Sec.~\ref{subsec:dysts-setup}. Figure~\ref{fig:excluded-example} illustrates the exclusion pattern on \emph{SprottA}, one of the quasi-periodic systems in the table.

\begin{table}[t]
\centering
\caption{Systems excluded from the cross-system analysis. For the top 19 systems, the recursive-baseline positive-$R^2$ horizon reaches the 40-Lyapunov-time threshold. The two \emph{zero-step baseline} systems have a recursive baseline whose $R^2$ fails at the first prediction step, so they have no defined positive-$R^2$ horizon. Regime labels summarize the sampled trajectories under the integration protocol of Sec.~\ref{subsec:data-setup}: \emph{quasi-periodic} systems have empirical Lyapunov exponent indistinguishable from zero; \emph{near-periodic} systems show autocorrelation peak $\geq 0.95$ with a small but nonzero empirical Lyapunov exponent; \emph{weakly} and \emph{slowly chaotic} systems retain genuine sensitive dependence, but the evaluation window is too short to resolve their recursive-baseline error growth.}
\label{tab:excluded-systems}
\begin{tabular}{ll}
\toprule
System & Regime under the \\ & integration protocol \\
\midrule
CaTwoPlusQuasiperiodic  & Quasi-periodic \\
CircadianRhythm         & Quasi-periodic \\
DequanLi                & Quasi-periodic \\
ForcedFitzHughNagumo    & Quasi-periodic \\
ForcedVanDerPol         & Quasi-periodic \\
SprottA                 & Quasi-periodic \\
Torus                   & Quasi-periodic \\
\midrule
CellularNeuralNetwork   & Near-periodic \\
ForcedBrusselator       & Near-periodic \\
HenonHeiles             & Near-periodic \\
IsothermalChemical      & Near-periodic \\
SprottI                 & Near-periodic \\
SprottO                 & Near-periodic \\
StickSlipOscillator     & Near-periodic \\
Tsucs2                  & Near-periodic \\
\midrule
SprottL                 & Weakly chaotic \\
\midrule
SprottJ                 & Slowly chaotic \\
SprottK                 & Slowly chaotic \\
SprottN                 & Slowly chaotic \\
\midrule
ArnoldWeb               & Zero-step baseline \\
PanXuZhou               & Zero-step baseline \\
\bottomrule
\end{tabular}
\end{table}

\clearpage
\renewcommand{\bibsection}{\section*{References}}
\bibliography{references}

\end{document}